\documentclass[10pt,twocolumn,letterpaper]{article}
\usepackage{cvpr}
\usepackage{eso-pic}
\usepackage{times}
\usepackage{graphicx}
\usepackage{amsmath}
\usepackage{amssymb}
\usepackage{cases}
\usepackage{setspace}
\usepackage{overpic}
\usepackage{rotating}
\graphicspath{{figures/}}
\usepackage[sort,numbers]{natbib}
\usepackage{subeqnarray}
\usepackage{multirow}
\usepackage{tabularx}
\usepackage{epstopdf}
\usepackage{subfigure}
\usepackage{enumerate}
\usepackage{color}
\usepackage{xcolor}

\makeatother

\usepackage[pagebackref=true,breaklinks=true,letterpaper=true,colorlinks,bookmarks=false]{hyperref}

\cvprfinalcopy % *** Uncomment this line for the final submission

\def\cvprPaperID{} % *** Enter the CVPR Paper ID here

\begin{document}
\vspace{-0.2cm}
%%%%%%%%% TITLE
\title{NTIRE 2020 Challenge on Perceptual Extreme
Super-Resolution:\\ Methods and Results}

\author{Kai Zhang
\and Shuhang Gu
\and Radu Timofte
%{OPPO}
\and Taizhang Shang
\and Qiuju Dai
\and Shengchen Zhu
\and Tong Yang
\and Yandong Guo
%{CIPLAB}
\and Younghyun Jo
\and Sejong Yang
\and Seon Joo Kim
%HiImageTeam}
\and Lin Zha
\and Jiande Jiang
\and Xinbo Gao
\and Wen Lu
%{ECNU}%Team Name
\and Jing Liu
%\subsection*{SIA}
\and Kwangjin Yoon
\and Taegyun Jeon
%TTI team}
\and Kazutoshi Akita
\and Takeru Ooba
\and Norimichi Ukita
%{DeepBlueAI}
\and Zhipeng Luo
\and Yuehan Yao
\and  Zhenyu Xu
%{APTX4869}
\and Dongliang He
\and Wenhao Wu
\and Yukang Ding
\and Chao Li
\and Fu Li
\and Shilei Wen
%{CNDP-Lab}
\and Jianwei Li
%{MSMers}
\and Fuzhi Yang
\and Huan Yang
\and Jianlong Fu
%{kaws}
\and Byung-Hoon Kim
\and JaeHyun Baek
\and Jong Chul Ye
%{UIUC-IFP}
\and Yuchen Fan
\and Thomas S. Huang
%{KU\_ISPLB}
\and Junyeop Lee
\and Bokyeung Lee
\and Jungki Min
\and Gwantae Kim
\and Kanghyu Lee
\and Jaihyun Park
% MsSrModel
\and Mykola Mykhailych
%{MoonCloud}
\and Haoyu Zhong
\and Yukai Shi
\and Xiaojun Yang
\and Zhijing Yang
\and Liang Lin
%{SuperT}
\and Tongtong Zhao
\and Jinjia Peng
\and Huibing Wang
%{sysu-AIR}
\and Zhi Jin
\and Jiahao Wu
\and Yifu Chen
\and Chenming Shang
\and Huanrong Zhang
%\subsection*{KU\_ISPL\_A}
\and Jeongki Min
%{CET\textunderscore CVLab}
\and Hrishikesh P S
\and Densen Puthussery
\and Jiji C V
}
%\vspace{-0.2cm}
\maketitle

\begin{abstract}
%\vspace{-0.2cm}
This paper reviews the NTIRE 2020 challenge on perceptual extreme
super-resolution with focus on proposed solutions and results.
The challenge task was to super-resolve an input image with a magnification factor $\times$16 based on a set of prior examples of low and corresponding high resolution images.
The goal is to obtain a network design capable to produce high resolution results with the best perceptual quality and similar to the ground truth. The track had 280 registered participants, and 19 teams submitted the final results. They gauge the state-of-the-art in single image super-resolution.
\end{abstract}

\vspace{-0.2cm}
\section{Introduction}
%\vspace{-0.1cm}
\let\thefootnote\relax\footnotetext{K. Zhang (kai.zhang@vision.ee.ethz.ch, ETH Zurich), S. Gu, and R. Timofte are the challenge organizers, while the other authors participated in the challenge. Appendix~\ref{sec:teams} contains the authors' teams and affiliations.
\\NTIRE webpage: \\\url{https://data.vision.ee.ethz.ch/cvl/ntire20/}}

Recent years have witnessed tremendous success of using deep neural networks (DNNs) to generate a high-resolution (HR) image from a low-dimensional input~\cite{Timofte_2018_CVPR_Workshops, Blau_2018_ECCV_Workshops,cai2019ntire,gu2019aim,lugmayr2019aim,zhang2019aim}.
On the one hand, DNNs-based single image super-resolution (SR) for bicubic degradation is continuously showing improvements in terms of PSNR and perceptual quality~\cite{Timofte-ACCV-2014,dong2015image,Timofte_2017_CVPR_Workshops, Timofte_2018_CVPR_Workshops,Blau_2018_ECCV_Workshops,ignatov2018pirm,gu2019aim,li2020group}. In particular, several fundamental conclusions have been drawn: (i) DNNs based SR with pixel-wise loss (such as L1 loss and L2 loss) tends to produce oversmoothed output for a large scale factor due to the pixel-wise average problem~\cite{ledig2017photo};
(ii) The perceptual quality of super-resolved image could be improved by using VGG perceptual loss and generative adversarial (GAN) loss~\cite{goodfellow2014generative,ledig2017photo,wang2018esrgan}; (iii) There is a trade-off between reconstruction accuracy and perceptual quality, which means no DNNs-based method can achieve its best PSNR and best perceptual quality at the same time~\cite{Blau_2018_ECCV_Workshops}.
While perceptual SR for bicubic degradation at a moderate scale factor (\eg, $\times$4) has achieved significant progress~\cite{ledig2017photo,wang2018esrgan,zhang2018rcan,zhang2019deep,zhang2020deep}, the case with an extremely large scale factor has received little attention~\cite{gu2019aim,bhler2020deepsee}. On the other hand, realistic HR image synthesis from a latent low-dimensional vector based on GAN has shown great success for natural image~\cite{brock2018large} and face image~\cite{karras2019style}.
However, how to effectively generate a perceptually pleasant HR image from a low-resolution (LR) image with a very large scale factor remains an open problem.

Jointly with NTIRE 2020 workshop we have an NTIRE challenge on perceptual extreme super-resolution, that is, the task of super-resolving an LR image to a perceptually pleasant HR image with a magnification factor $\times$16. Although AIM 2019 extreme SR challenge~\cite{gu2019aim} has
considered the fidelity track and perceptual track, it has been concluded that: (i) the PSNR-orientated methods consistently give rise to oversmoothed results; (ii) there still remains a large room for perceptual quality improvement.
As a result, this challenge only has one track which aims to seek effective solutions for perceptual extreme SR.

\clearpage
\section{NTIRE 2020 Challenge}
This challenge is one of the NTIRE 2020 associated challenges on: deblurring~\cite{nah2020ntire}, nonhomogeneous dehazing~\cite{ancuti2020ntire}, perceptual extreme super-resolution~\cite{zhang2020ntire}, video quality mapping~\cite{fuoli2020ntire}, real image denoising~\cite{abdelhamed2020ntire}, real-world super-resolution~\cite{lugmayr2020ntire}, spectral reconstruction from RGB image~\cite{arad2020ntire} and demoireing~\cite{yuan2020demoireing}.

The objectives of the NTIRE 2020 challenge on perceptual extreme super-resolution challenge are:
(i) to advance researches on perceptual SR at an extremely large scale factor; (ii) to compare the effectiveness of different methods and (iii) to offer an opportunity for academic and industrial attendees to interact and explore collaborations.

\subsection{DIV8K Dataset~\cite{gu2019div8k}}
Following~\cite{gu2019div8k}, the DIV8K dataset which contains 1,700 DIVerse 8K resolution RGB images is employed in this challenge. The HR DIV8K is divided into 1,500 training images, 100 validation images and 100 testing images.
The corresponding LR images in this challenge is obtained via default
setting (bicubic interpolation) of Matlab function \texttt{imresize} with scale factor 16. The testing HR images are completely hidden from the participants during the whole challenge. In order to get access to the data and submit the testing HR results, registration on Codalab (\url{https://competitions.codalab.org/}) is required.

\subsection{Track and Competition}

\noindent{\textbf{Track }}
This challenge has only one track. The aim is to obtain a network design capable to produce high resolution results with the best perceptual quality and similar to the ground truth.

\vspace{0.07cm}
\noindent{\textbf{Challenge phases }}

\textit{(1) Development phase:} the participants got the 1,500 HR training images and 100 LR validation images of the DIV8K dataset; the participants got the LR training images via Matlab's \texttt{imresize} function.
Due to the storage constraints, the participants uploaded the center 1,000$\times$1,000 HR validation results to an online validation server to get immediate feedback.
During this phase, the provided feedback consisted from Peak Signal-to-Noise Ratio (PSNR) and Structural Similarity Index (SSIM)~\cite{Wang-TIP-2004} results. However, since both PSNR and SSIM are not suitable for perceptual ranking, the validation results are only used to test whether the cropped images are correctly uploaded.
\textit{(2) Testing phase:} the participants got 100 LR testing images; the participants submitted their center 1,000$\times$1,000 results of the super-resolved HR images to Codalab and emailed the code and factsheet to the organizers; the participants got the final results after the end of the challenge.

\vspace{0.07cm}
\noindent{\textbf{Evaluation protocol }}
Apart from PSNR and SSIM, the quantitative measures also includes LPIPS (Learned Perceptual Image Patch Similarity)~\cite{zhang2018unreasonable} and no-reference PI (Perceptual Index)~\cite{ma2017learning,Blau_2018_ECCV_Workshops} which have been acknowledged as useful perceptual metrics. The evaluation is performed on the center 1,000$\times$1,000 HR results for convenience and consistency with the reported results on the challenge servers. The final ranking employed also a user study. In order to have a thorough evaluation, the self-reported number of parameters and running time per testing image are also reported.

\section{Challenge Results}
\begin{figure*}[!htbp]\footnotesize
	%\centering
\hspace{-0.22cm}
\begin{tabular}{c@{\extracolsep{0.00em}}c@{\extracolsep{0.00em}}c@{\extracolsep{0.00em}}c@{\extracolsep{0.00em}}c}

        \includegraphics[width=0.19\textwidth]{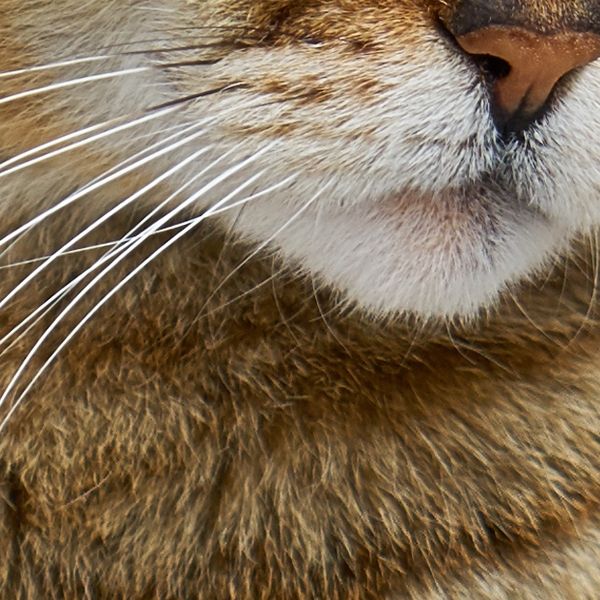}~
		&\includegraphics[width=0.19\textwidth]{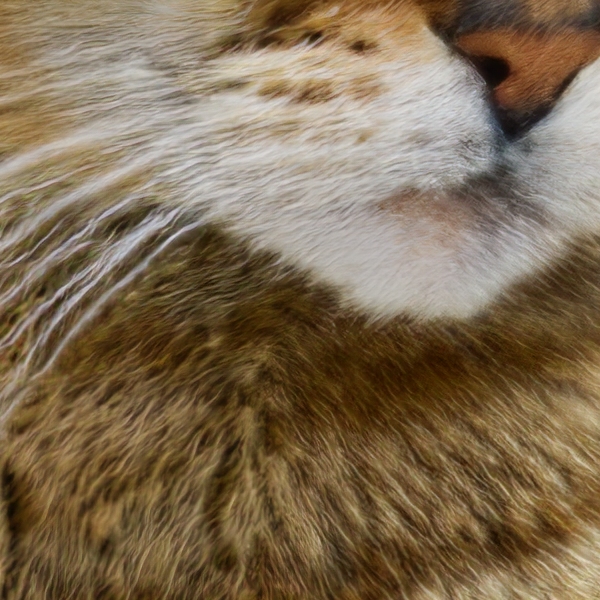}~
		&\includegraphics[width=0.19\textwidth]{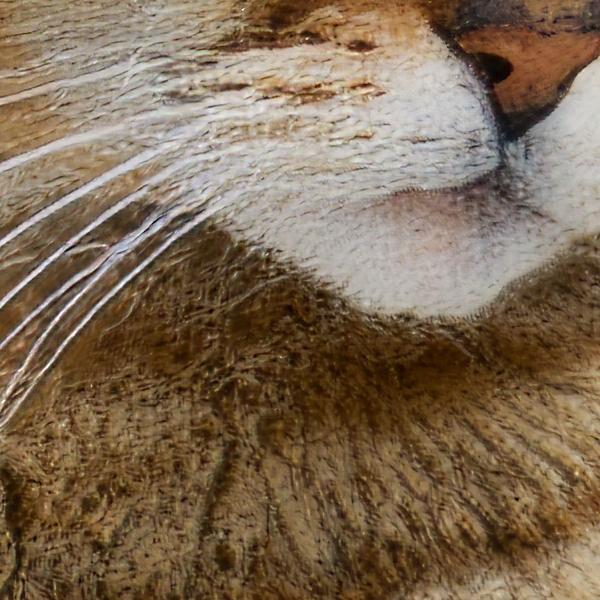}~
        &\includegraphics[width=0.19\textwidth]{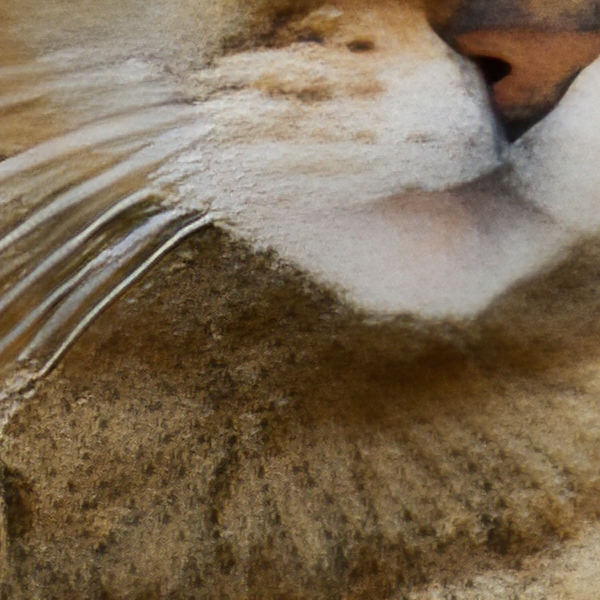}~
        &\includegraphics[width=0.19\textwidth]{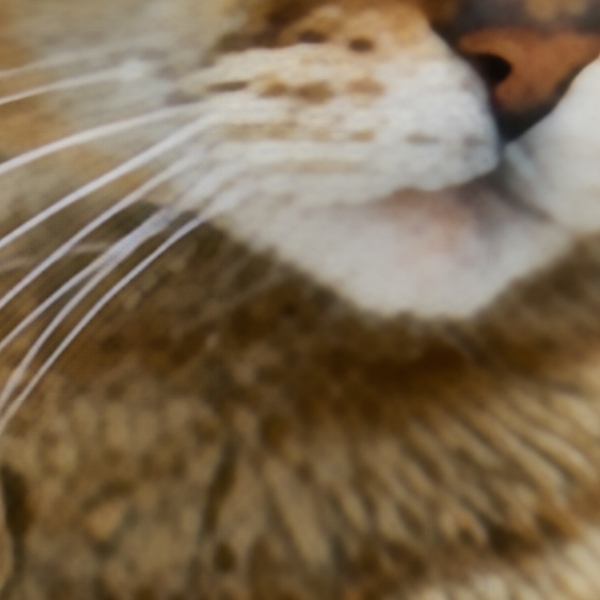}~\\

 PSNR/SSIM/LPIPS & 20.42/0.3270/0.426 & 20.31/0.3544/0.441 &20.90/0.3980/0.561  &  22.14/0.4904/0.657 \\
 (a) HR & (b) OPPO-Research &(c) CIPLAB & (d) HiImageTeam & (e) ECNU  \\

        \includegraphics[width=0.19\textwidth]{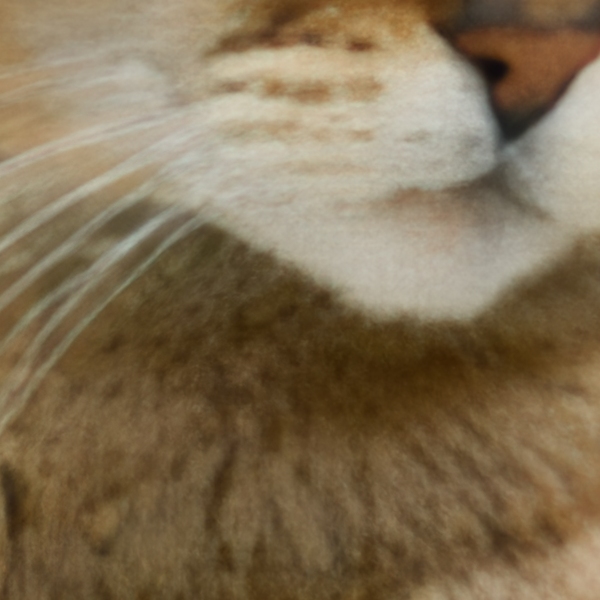}~
		&\includegraphics[width=0.19\textwidth]{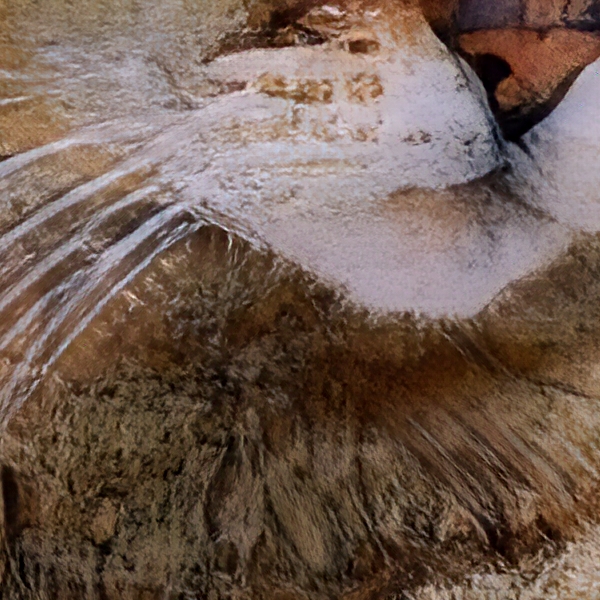}~
		&\includegraphics[width=0.19\textwidth]{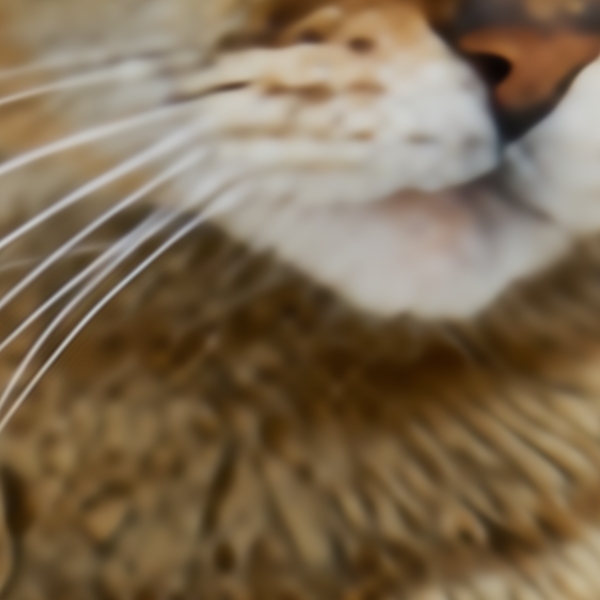}~
        &\includegraphics[width=0.19\textwidth]{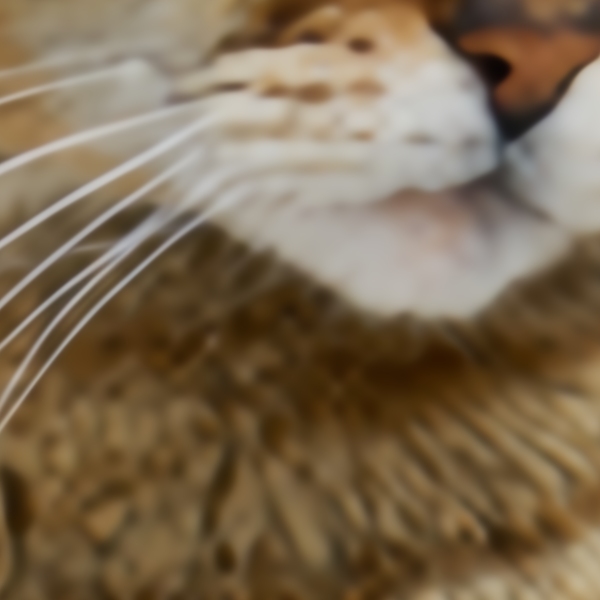}~
        &\includegraphics[width=0.19\textwidth]{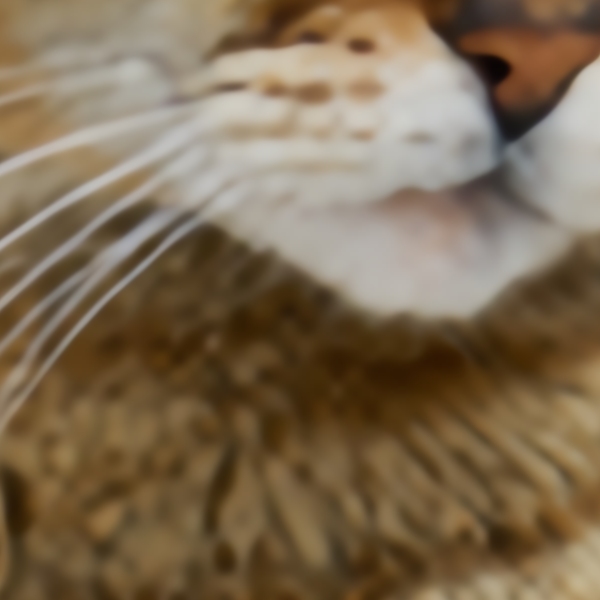}~\\
20.92/0.4562/0.602 & 17.98/0.2875/0.505 & 22.26/0.5965/0.714 & 22.20/0.4947/0.712 & 21.88/0.4826/0.740  \\
(f) SIA & (g) TTI & (h) DeepBlueAI & (i) APTX4869 & (j) CNDP-Lab  \\
	\end{tabular}
    \vspace{0.2cm}
	\caption{SR results (600$\times$600 crop) by different methods.}
	\label{fig:output_vr1}
	\vspace{-0.0cm}
\end{figure*}

\begin{figure*}[!htbp]\footnotesize
	%\centering
\hspace{-0.22cm}
\begin{tabular}{c@{\extracolsep{0.00em}}c@{\extracolsep{0.00em}}c@{\extracolsep{0.00em}}c@{\extracolsep{0.00em}}c}

        \includegraphics[width=0.19\textwidth]{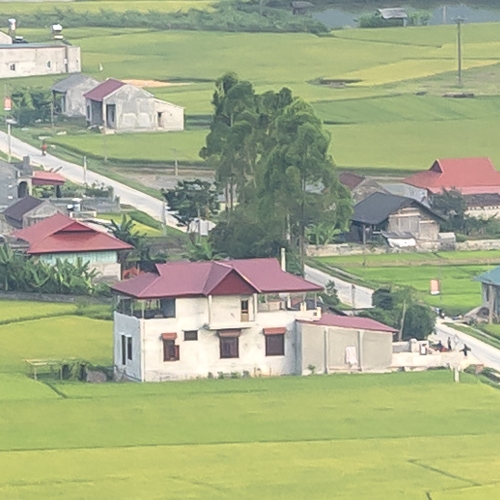}~
		&\includegraphics[width=0.19\textwidth]{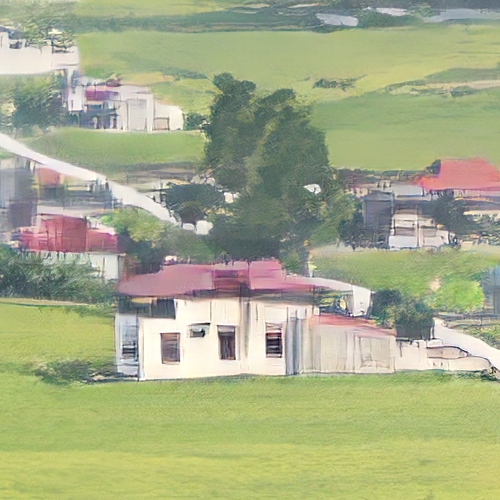}~
		&\includegraphics[width=0.19\textwidth]{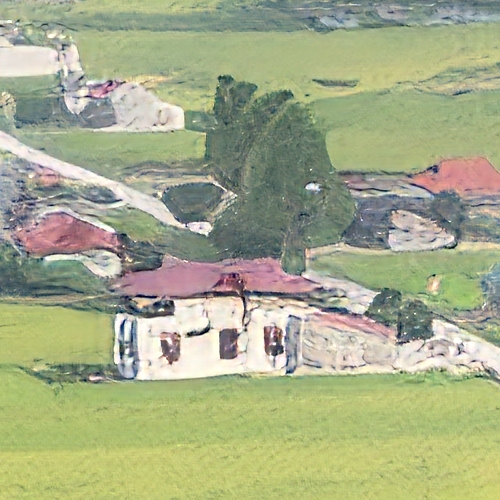}~
        &\includegraphics[width=0.19\textwidth]{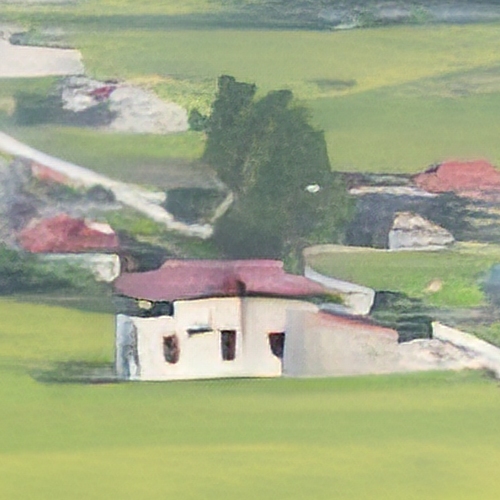}~
        &\includegraphics[width=0.19\textwidth]{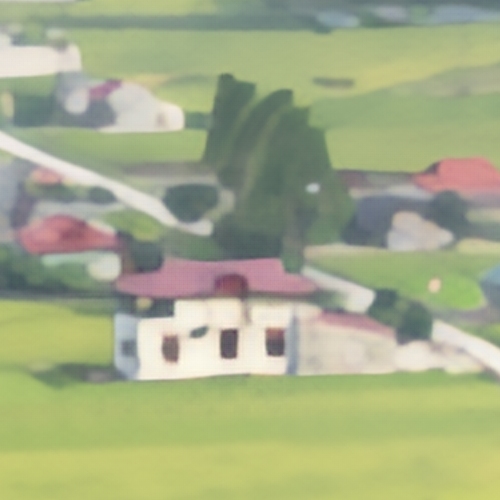}~\\

 PSNR/SSIM/FPIPS & 19.98/0.4441/0.367 & 19.43/0.4273/0.370 &20.30/0.4990/0.410  &  22.14/0.5753/0.547 \\
 (a) HR & (b) OPPO-Research &(c) CIPLAB & (d) HiImageTeam & (e) ECNU  \\

        \includegraphics[width=0.19\textwidth]{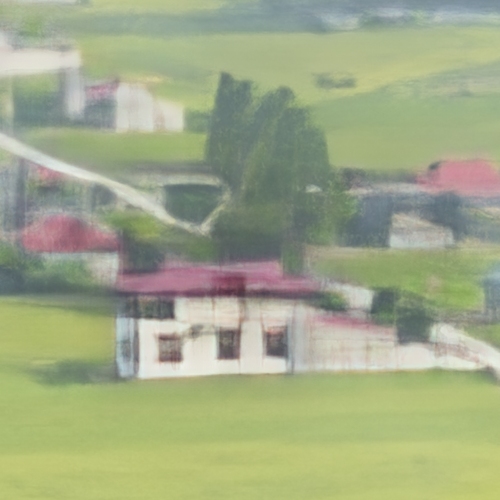}~
		&\includegraphics[width=0.19\textwidth]{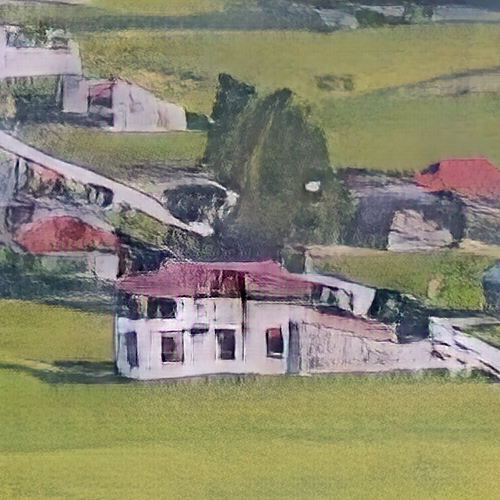}~
		&\includegraphics[width=0.19\textwidth]{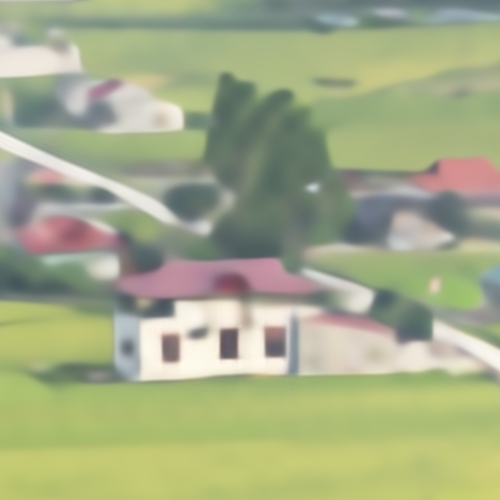}~
        &\includegraphics[width=0.19\textwidth]{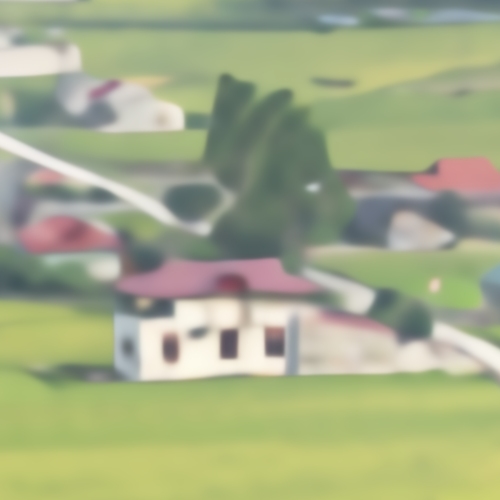}~
        &\includegraphics[width=0.19\textwidth]{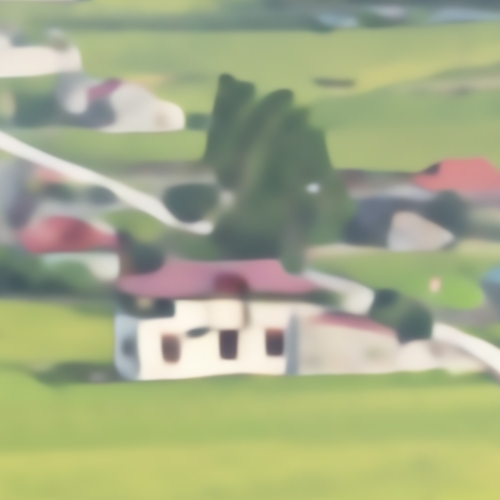}~\\
20.04/0.5328/0.486 & 15.56/0.3910/0.421 & 22.27/0.5812/0.616 & 22.27/0.5796/0.626 & 22.15/0.5769/0.621  \\
(f) SIA & (g) TTI & (h) DeepBlueAI & (i) APTX4869 & (j) CNDP-Lab  \\
	\end{tabular}
    \vspace{0.2cm}
	\caption{SR results (500$\times$500 crop) by different methods.}
	\label{fig:output_vr2}
	\vspace{-0.0cm}
\end{figure*}

From 280 registered participants, 19 teams entered in the final phase and submitted results, codes, and factsheets. Table~\ref{table_track1} reports the final test results and rankings of the challenge. Note that the methods trained with GAN loss are grouped together, and only the best six methods are ranked by user study.
The results of winner teams in AIM 2019 extreme SR challenge are also reported for comparison.
Figures~\ref{fig:output_vr1} and~\ref{fig:output_vr2} show the visual results and associated PSNR, SSIM, LPIPS and PI values of different methods.

From Table~\ref{table_track1} and Figures~\ref{fig:output_vr1} and~\ref{fig:output_vr2}, we can have the following observations.
First, the OPPO-Research team is the first place winner of this challenge, while CIPLAB and HiImageTeam win the second place and third place, respectively.
Second, among the top-6 methods, ECNU achieves a good trade-off between reconstruction accuracy and perceptual quality, while HiImageTeam shows the best trade-off between number of parameters and inference time.
Third, DeepBlueAI achieves the best PSNR performance, however, it fails to generate results with competitive perceptual quality.
Fourth, LPIPS and PI are relatively reliable perceptual measures in comparison with PSNR and SSIM for perceptual extreme SR.
Fifth, it is easy to distinguish the ground-truth HR images from the super-resolved HR images.

\begin{table*}[!t]
\caption{Results of NTIRE 2020 perceptual extreme SR challenge. The PSNR, SSIM~\cite{Wang-TIP-2004}, LPIPS~\cite{zhang2018unreasonable} and PI~\cite{Blau_2018_ECCV_Workshops} are calculated on the center 1,000$\times$1,000 subimages of the DIV8K test images.
}
\centering
\resizebox{\linewidth}{!}
{
\begin{tabular}{ll|llll|rrccccc<{\centering}}

 \multirow{2}{*}{Team} &  \multirow{2}{*}{Author}  &   \multirow{2}{*}{PSNR~$\uparrow$}&   \multirow{2}{*}{SSIM~$\uparrow$}&   \multirow{2}{*}{LPIPS~$\downarrow$}&\multirow{2}{*}{PI~$\downarrow$}  & \#Params  & Time &  \multirow{2}{*}{Platf.} &  \multirow{2}{*}{Ens.} & \multirow{2}{*}{GPU} & Extra &\multirow{2}{*}{Loss}  \\%\cline{4-10}

    &   &   &  &  &   &  [M]  & [s] &  &   &  &  data       \\\hline\hline
\\
\multicolumn{13}{c}{\large Methods optimized with GAN losses}
\\
\hline
%1
OPPO-Research &sss& 23.38$_{(12)}$ & 0.5504$_{(15)}$  & 0.348$_{(1)}$&3.97$_{(2)}$	&  20.5  &	8.1 &PyTorch& Model &V100	& DF2K, OST & L1, VGG-P, GAN\\
%2
CIPLAB & heyday097	& 22.77$_{(15)}$ & 0.5251$_{(16)}$  & 0.352$_{(2)}$&3.76$_{(1)}$  & 33.0   & 3.0	 & PyTorch	& None   & Xp	& None & Huber, LPIPS, FM, GAN\\
%3
HiImageTeam & HiImageTeam& 23.53$_{(11)}$&0.5624$_{(13)}$& 0.368$_{(3)}$& 4.38$_{(4)}$ &  4.0 & 1.0	 &PyTorch& None   & RTX	& None&L1, VGG-P, GAN\\

\textit{Winner AIM19~\cite{gu2019aim}} & \textit{BOE-IOT-AIBD}& 24.52 & 0.5800 & 0.418& 6.28& - & 47.1 & PyTorch & None & Titan X & None &   L1, VGG-P, GAN     \\

%4
ECNU &lj1995& 25.56$_{(4)}$  & 0.6336$_{(6)}$  &  0.497$_{(6)}$&	8.10$_{(8)}$&   57.9 &26.0 &PyTorch& Self  &	1080Ti	&None&L1, VGG-P, GAN \\
%5
SIA & yoon28&22.86$_{(14)}$&0.5896$_{(11)}$ & 0.434$_{(5)}$& 5.81$_{(6)}$ & 16.0 & 360.0 & PyTorch	& Self  & CPU & None & L1, VGG-P, GAN \\

%6
TTI & iim-nike	& 19.16$_{(17)}$ & 0.4993$_{(17)}$ & 0.377$_{(4)}$& 3.99$_{(3)}$& 26.5 & 19.5 & PyTorch & None & V100 & None & L1, VGG-P, GAN\\
%17
%perceptual bad
sysu-AIR & Zhi\_Jin\_SYSU&23.94$_{(10)}$&0.5545$_{(14)}$ & 0.510$_{(7)}$&4.99$_{(5)}$ & 2.1 & 4.3 & PyTorch & None & 2080Ti & None & L1, TV, FS, VGG-P, GAN\\

CET\_CVLab &hrishikeshps& 19.68$_{(16)}$&0.4290$_{(18)}$& 0.705$_{(19)}$ &7.46$_{(7)}$	&223.8    &440.0	 &TensorFlow	&None & CPU    &None &  L1, VGG-P, GAN \\

\hline
\\
\multicolumn{13}{c}{\large From here down are mostly L1/L2 optimized methods}\\
\hline

DeepBlueAI &DeepBlueAI& 25.70$_{(1)}$ & 0.6390$_{(2)}$ &  0.555$_{(9)}$&9.15$_{(9)}$	&  63.5  & 0.8	 & PyTorch	& Model &V100& None& L1; L2 \\

\textit{Winner AIM19~\cite{gu2019aim}} & \textit{NUAA-404} &25.63 & 0.6394 & 0.554 &9.21 & - & 30.0 & PyTorch & Self & 2080Ti & - & L1
\\
%8
APTX4869 &APTX4869	 & 25.62$_{(2)}$  &  0.6393$_{(1)}$ & 0.558$_{(11)}$& 9.25$_{(11)}$&20.8    &40.0	 &PyTorch	&Self    &V100	&DIV2K & L2, VGG-P, Style, GAN; L1 \\
%9
CNDP-Lab & albertron& 25.58$_{(3)}$  & 0.6367$_{(3)}$  & 0.556$_{(10)}$ &9.36$_{(14)}$ & 322.8 & 350.0& PyTorch & Self & RTX & None & L1\\
%10
MSMers &huayan	 & 25.56$_{(4)}$  & 0.6366$_{(4)}$  & 0.564$_{(12)}$& 9.40$_{(16)}$&16.0    &16.0	 &PyTorch	&Self &V100	&None & L1\\
%11
kaws & egyptdj	 & 25.43$_{(5)}$&0.6339$_{(5)}$& 0.568$_{(13)}$&9.37$_{(15)}$ & 126.2 & 9.76  & TensorFlow	& Self & V100 & None & L1 \\
%12
UIUC-IFP& fyc0624&  25.28$_{(7)}$ & 0.6339$_{(5)}$  & 0.553$_{(8)}$&9.28$_{(12)}$  &100.6& 38.0 &PyTorch &Self &1080Ti &None & L1\\
%13
KU\_ISPLB & givenjiang	&25.33$_{(6)}$ &0.6299$_{(7)}$& 0.582$_{(14)}$& 9.44$_{(17)}$ &1.9  & 70.0	 & Pytorch	& Self  &	RTX	& None & L1\\
%14
MsSrModel& nickolay	 &25.23$_{(8)}$ & 0.6259$_{(8)}$  & 0.586$_{(15)}$&9.29$_{(13)}$ &9.0 &1.3 &TensorFlow &Self& V100 & None & L2, VGG-P\\
%15
MoonCloud &pigfather315	 & 25.17$_{(9)}$  & 0.6250$_{(9)}$  & 0.587$_{(16)}$&9.23$_{(10)}$ &4.9    &3.2	 &PyTorch	&None    &V100 &None & L1\\
%16
SuperT &tongtong& 23.94$_{(10)}$&0.6060$_{(10)}$& 0.630$_{(17)}$&9.40$_{(16)}$ & 0.4 & 0.6 & Tensorflow & Self & V100 & DIV2K  & L1\\

%18
KU\_ISPL\_A & jkm\_ispl	 & 23.05$_{(13)}$ &0.5787$_{(12)}$& 0.667$_{(18)}$&11.4$_{(18)}$ & 0.3 &	30.0 & PyTorch	& None & 2080Ti & None & L1\\\hline
%19

\textit{Baseline} & \textit{Bicubic} & 24.22 & 0.6017 & 0.683&11.1

\end{tabular}
}
\label{table_track1}
\end{table*}

\vspace{0.08cm}
\noindent{\textbf{Architectures, losses and main ideas }}
All the proposed methods utilize deep neural networks for perceptual extreme SR. Overall, there are two import factors to improve the perceptual quality of super-resolved images, \ie, network architecture and loss function. For the network architecture, several teams, such as CIPLAB, APTX4869 and MSMers, proposed to extend existing state-of-the-art SR methods with a progressive upscaling strategy. Several other teams, such as OPPO-Research, SIA and DeepBlueAI, achieved a scale factor of 16 by directly modifying the upscaling layer. For the loss function, most of the teams adopted either L1 loss or the same loss (\ie, a combination of L1 loss, VGG perceptual loss~\cite{simonyan2014very,johnson2016perceptual} and relativistic GAN loss~\cite{jolicoeur2018relativistic}) proposed in ESRGAN~\cite{wang2018esrgan} as their final loss. In particular, CIPLAB replaced the VGG perceptual loss with LPIPS loss, and both CIPLAB and TTI adopted an U-Net~\cite{ronneberger2015u}-like discriminator~\cite{schonfeld2020u} for better local and global perceptual quality enhancement.

\vspace{0.08cm}
\noindent{\textbf{Ensembles }}
Most of the teams adopted commonly-used model-ensemble or self-ensemble~\cite{Timofte-CVPR-2016} to enhance the performance.

\vspace{0.08cm}
\noindent{\textbf{Train data }}
Most of the teams only used the provided DIV8K dataset~\cite{gu2019div8k} for training. OPPO-Research further adopted DIV2K~\cite{Agustsson_2017_CVPR_Workshops}, Flickr2K~\cite{Timofte_2017_CVPR_Workshops} and OST~\cite{wang2018recovering} datasets, while APTX4869 and SuperT used DIV2K~\cite{Agustsson_2017_CVPR_Workshops} as additional training data.

\vspace{0.08cm}
\noindent{\textbf{Conclusions }}
From the above analysis of different solutions, we can have several conclusions.
(i) The proposed methods improve the state-of-the-art for extreme SR. On one hand, compared to the best method proposed by NUAA-404  in AIM 2019 fidelity extreme SR challenge~\cite{gu2019aim}, DeepBlueAI achieved an average PSNR gain of 0.07dB.
On the other hand, compared to the best method proposed by BOE-IOT-AIBD in AIM 2019 perceptual extreme SR challenge~\cite{gu2019aim}, OPPO-Research, CIPLAB and HiImageTeam produced perceptually better results and an improved LPIPS value.
(ii) The perceptual extreme SR is far from being solved. Advanced training strategies, network architectures and perceptual losses for a good trade-off between reconstruction accuracy and perceptual quality require further study.
(iii) Due to the high ill-poseness of $\times$16 SR, DIV8K might not be sufficient to capture the diversity of natural images.

%------------------------------------------------------------------------
\section{Challenge Methods and Teams}
\label{sec:methods_and_teams}

\subsection*{OPPO-Research}
\begin{figure}[!h]
    \centering
    \subfigure[The network architecture of RFB-SRGAN.]{%
    \includegraphics[width=0.8\linewidth]{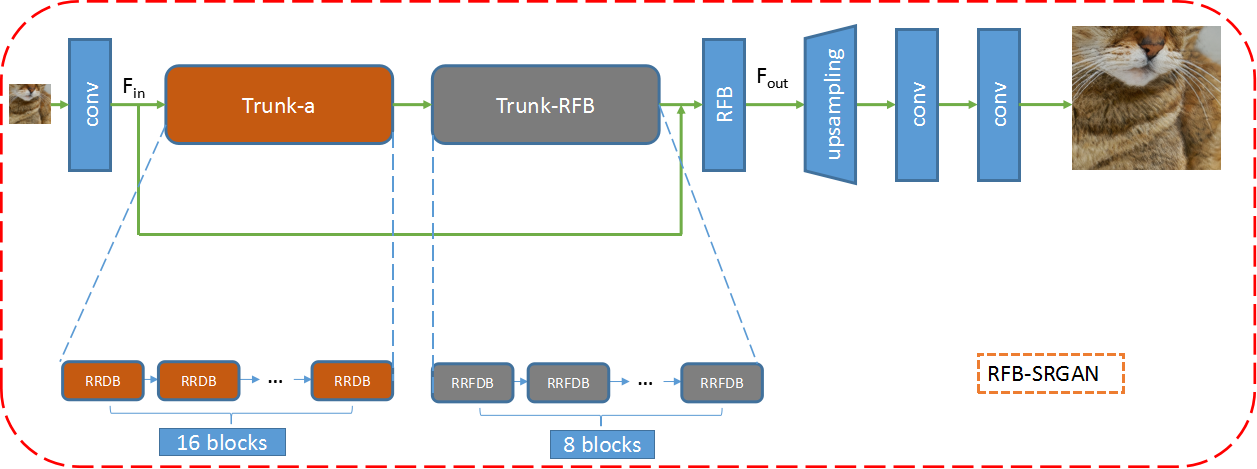}
    }
    \subfigure[RRDB block.]{%
    \includegraphics[width=0.4\linewidth]{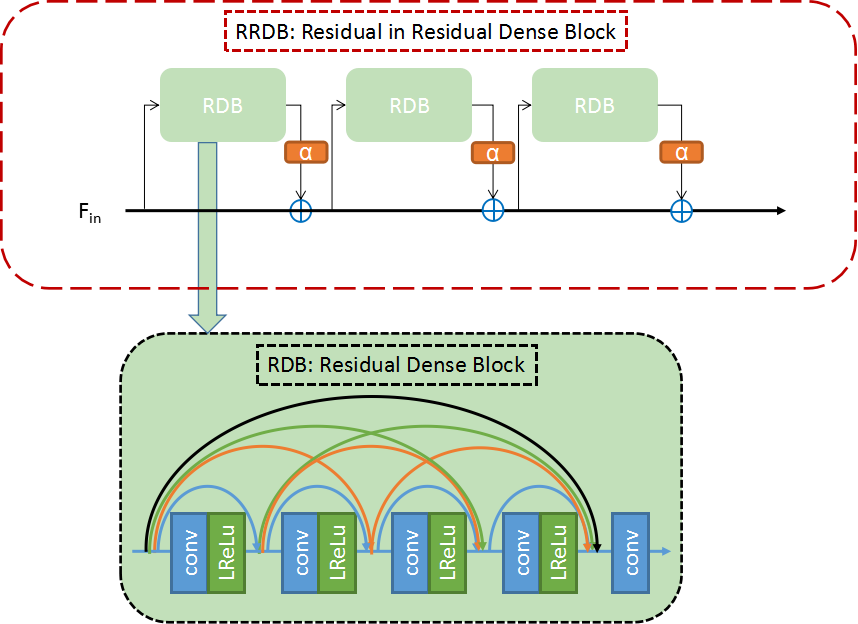}
    }
    \subfigure[RRFDB block.]{%
    \includegraphics[width=0.4\linewidth]{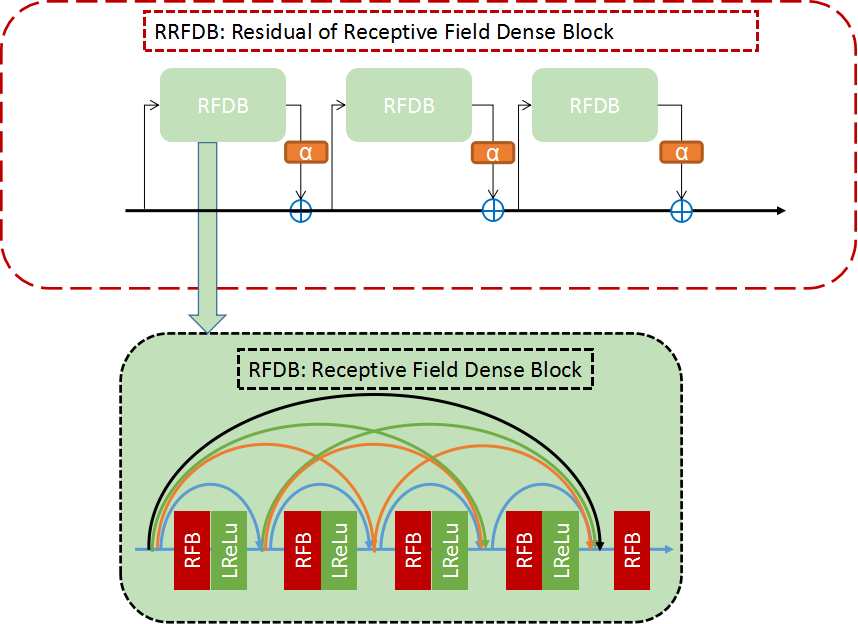}
    }
    %\vspace{-1mm}
    \caption{OPPO-Research's network architecture.}
    \label{fig:oppo}
\end{figure}

OPPO-Research proposed a novel Super-Resolution GAN~\cite{shang2020rfbesrgan}, namely RFB-SRGAN, based on ESRGAN~\cite{wang2018esrgan}.
As shown in Figure~\ref{fig:oppo}, RFB-SRGAN consists of 5 parts, including the feature extraction module, the Trunk-a module, the Trunk-RFB module, the features up-sampling module, and the final convolution reconstruction module.
Specifically, the feature extraction module is composed of a convolution layer.
The Trunk-a module consists of 16 Residual in Residual Dense Blocks (RRDBs). The Trunk-RFB module is stacked of 8 Residual of Receptive Field Dense Blocks (RRFDBs), and each RRFDB contains 5 Receptive Fields Blocks (RFBs)~\cite{liu2018receptive}.
The feature up-sampling module uses sub-pixel convolution~\cite{shi2016real} and nearest neighborhood interpolation which can greatly reduce the time cost while maintaining satisfactory performance.
The final convolution reconstruction module consists of two layers of convolutions.

The training process is divided into two stages. In the first stage, a PSNR-oriented model was trained with L1 loss. The learning rate is initialized as $2\times10^{-4}$ and decayed by a factor of $2$ every $2\times10^5$ steps. In the second stage, the generator was initialized by the pre-trained PSNR-oriented model and further trained with the loss function in ESRGAN~\cite{wang2018esrgan}. The training mini-batch size is set to $16$.
The resolution of the cropped HR images is 512$\times$512.
The performance of models at different iteration stages are scored, then the top-scored models are fused to obtain the final model.

%\vspace{-0.2cm}

\subsection*{CIPLAB}
\begin{figure}[!htbp]
\small
\centering
\includegraphics[width=\linewidth]{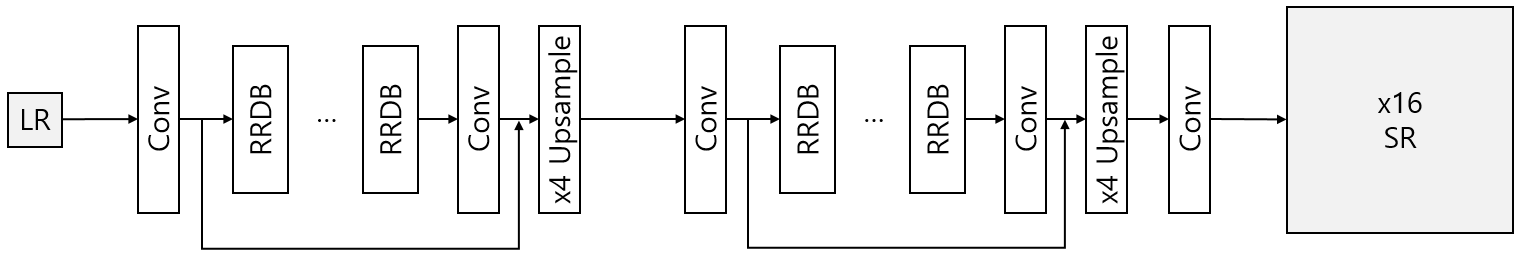}
\caption{CIPLAB's generator network.}
\label{fig:generator}
\end{figure}

CIPLAB proposed to use GAN~\cite{goodfellow2014generative} with LPIPS~\cite{zhang2018unreasonable} loss for perceptual extreme SR~\cite{jo2020investigating} instead of using GAN with VGG perceptual loss~\cite{johnson2016perceptual}.
Generally, the loss functions for perceptual SR is the adversarial loss \cite{goodfellow2014generative} with the VGG perceptual loss \cite{johnson2016perceptual}.
Such a loss combination has worked well for $\times$4 SR, however, it is found that it is not work well for $\times$16 SR due to highly hallucinated noise and less precise details.
Because VGG network is trained for image classification, it may not the best choice for the SR task.
On the other hand, the learned perceptual similarity (LPIPS)~\cite{zhang2018unreasonable} is trained with a dataset of human perceptual similarity judgments, thus it is expected to be a more proper choice for perceptual SR. For this reason, LPIPS is adopted instead of the VGG perceptual loss.
In addition, the discriminator's feature matching loss helps to increase the quality of the results, and Huber loss prevents color permutation.
The proposed generator, as shown in Figure~\ref{fig:generator}, consists of two ESRGAN generators for $\times$16 SR.
For the discriminator, an U-Net network as in~\cite{schonfeld2020u} is adopted to  judge real and fake for the compressed space from the encoder head and every pixel from the decoder head.
Such a discriminator allows to provide detailed per-pixel feedback to the generator while maintaining the global context.
It is empirically found that the discriminator can recover more details than normal encoder structure discriminator.

In the training, the HR patch size is set to 384$\times$384, and the corresponding LR patch size is 24$\times$24.
For both generator and discriminator, Adam optimizer with learning rate $1\times10^{-5}$ is adopted.
The generator is first trained with L2 loss and mini-batch size 3 for 50K iterations. Then the model is trained with the proposed new combination of different loss functions and mini-batch size 2 for about 60K iterations.

\vspace{-0.2cm}

\begin{figure}[!htbp]
\small
\centering
\includegraphics[width=\linewidth]{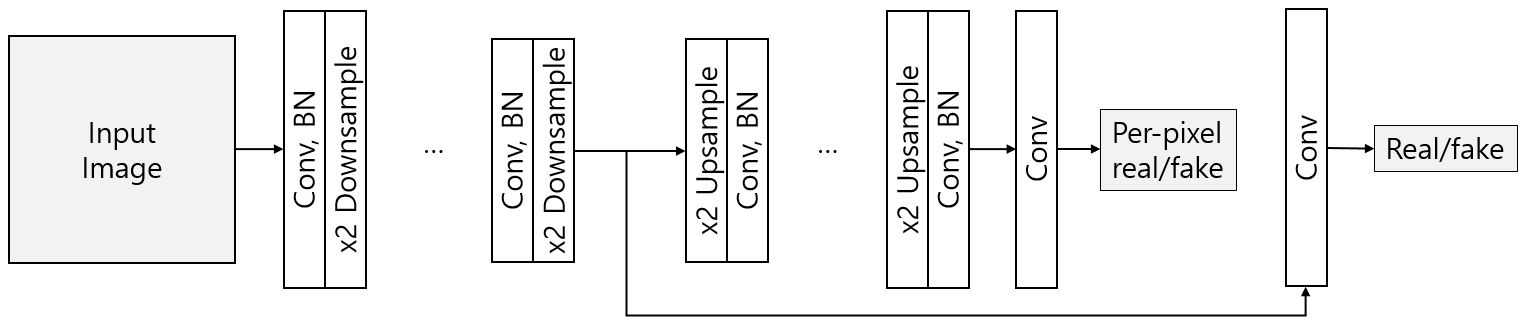}
\caption{CIPLAB's discriminator network.}
\label{fig:discriminator}
\end{figure}
\vspace{-0.2cm}

\subsection*{HiImageTeam}
\begin{figure}[!h]
    \centering
    \includegraphics[width=\linewidth]{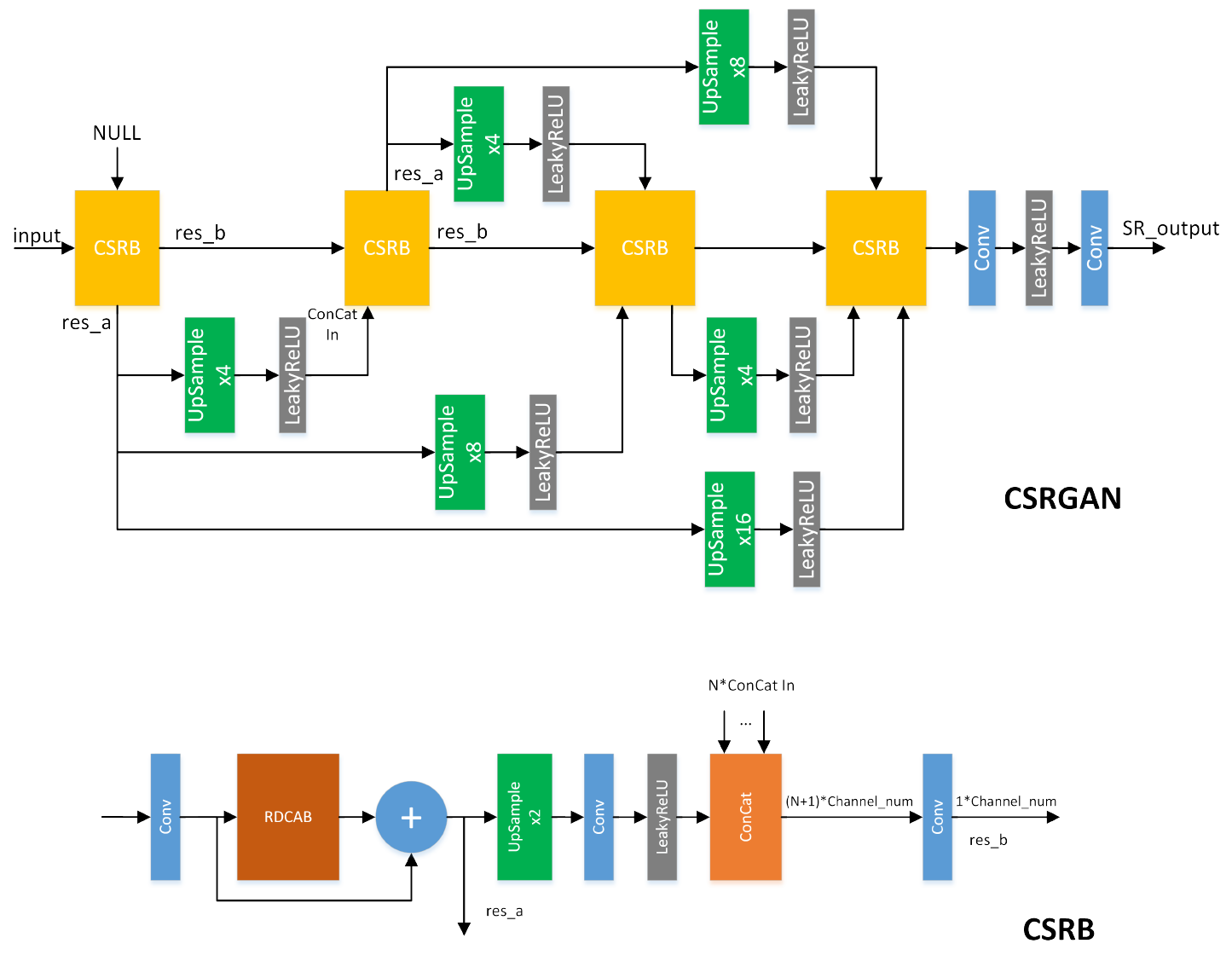}
    %\vspace{-1mm}
    \caption{HiImageTeam's network architecture.}
    \label{fig:HiImageTeam}
\end{figure}

\begin{figure}[!h]
    \centering
    \includegraphics[width=\linewidth]{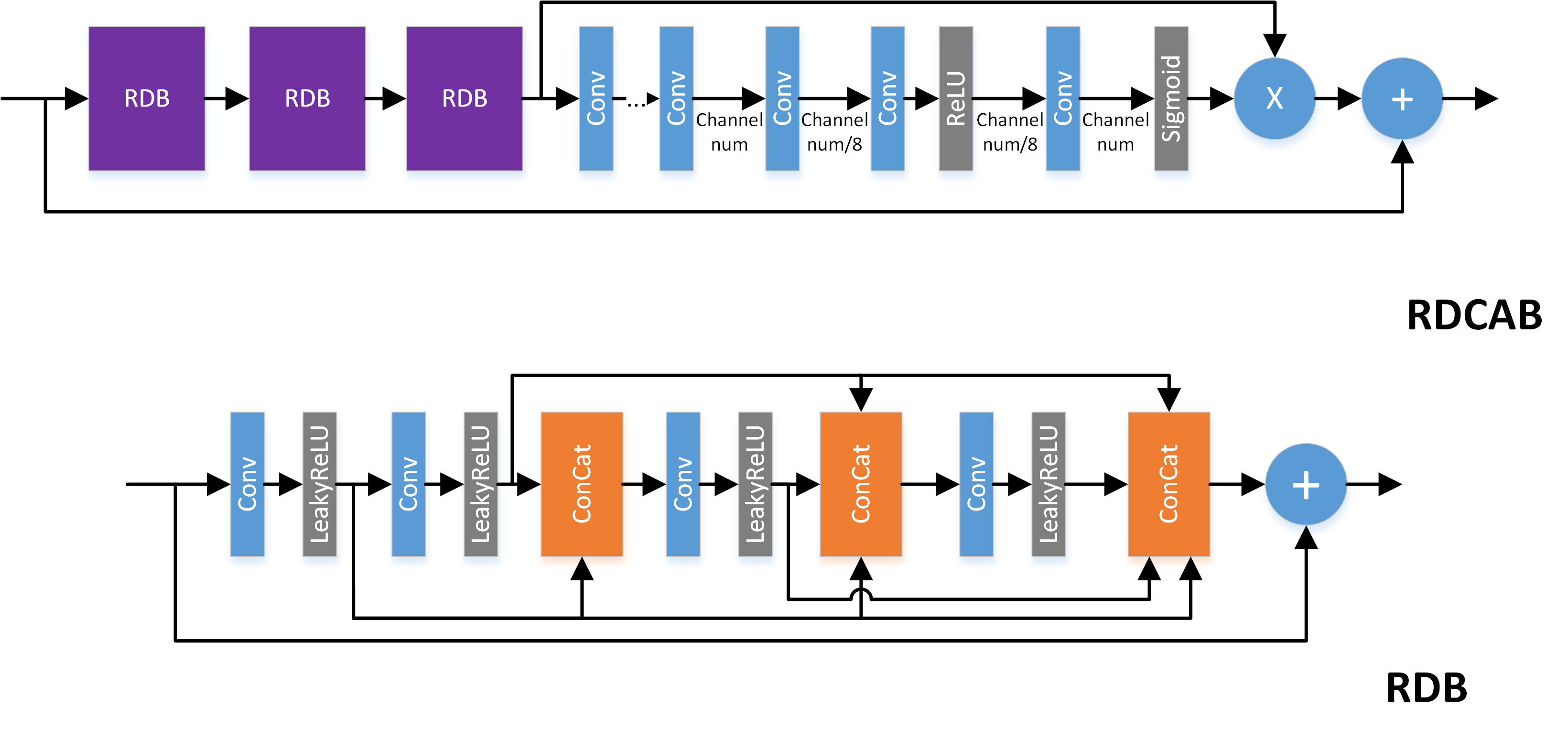}
    %\vspace{-1mm}
    \caption{HiImageTeam's Residual Dense Channel Attention Block.}
    \label{fig:HiImageTeam_fig2}
\end{figure}

HiImageTeam proposed Cascade SR-GAN (CSRGAN) for perceptual extreme SR. As shown in Figure~\ref{fig:HiImageTeam}, CSRGAN achieves an upscaling of $\times$16 via four successive $\times$2 subnetworks (CSRB). In order to improve the performance, a novel residual dense channel attention block (see Figure~\ref{fig:HiImageTeam_fig2}) is proposed. Final CSRGAN uses VGG perceptual loss and GAN loss to enhance the perceptual quality of super-resolved images.

\subsection*{ECNU}

\begin{figure}[!htb]
	\centering
	\includegraphics[width=\linewidth]{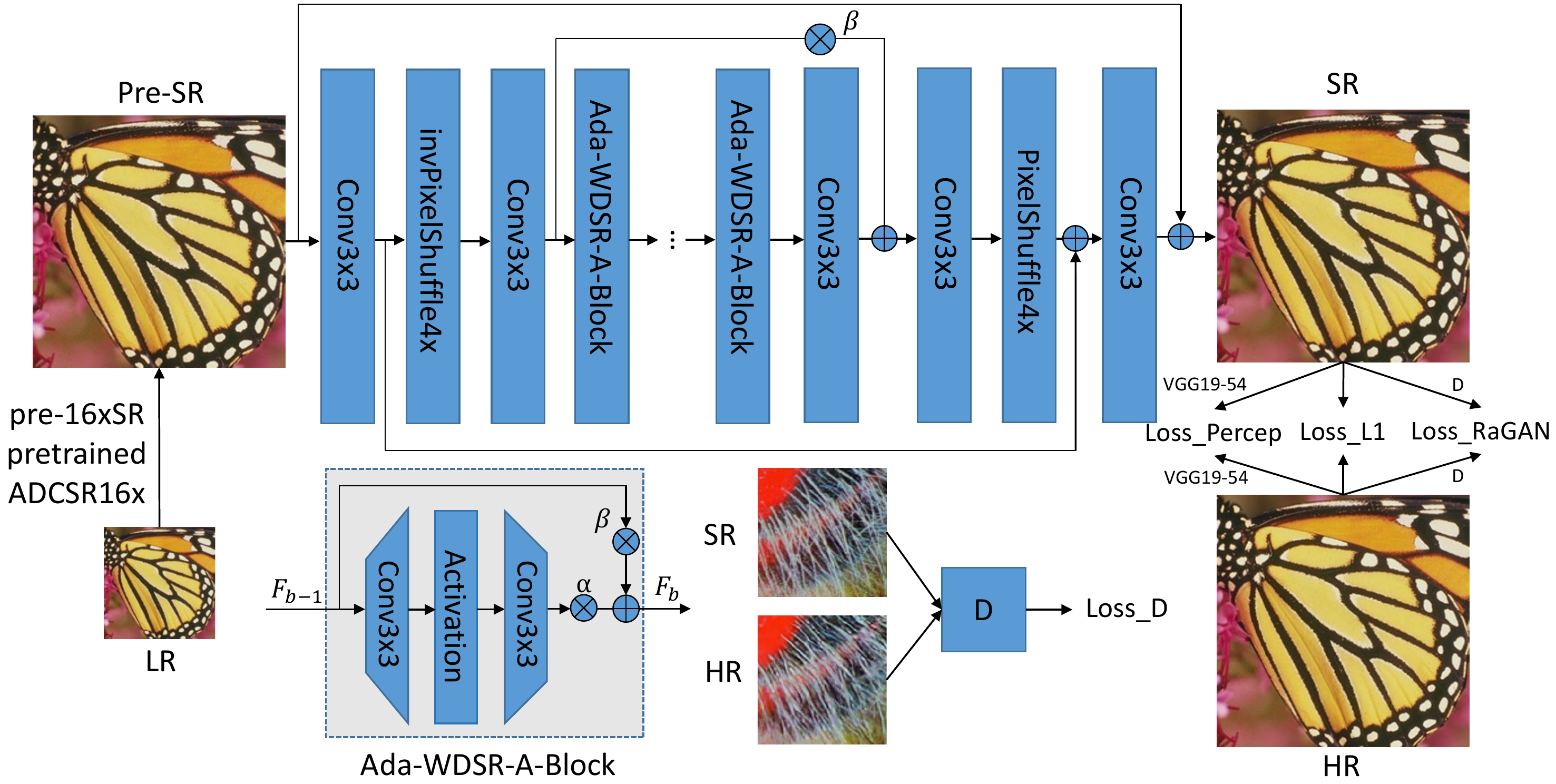}\\
	\caption{ECNU's network architecture. }
	\label{fig_ecnu}
\end{figure}

ECNU proposed a Two Stages Super Resolution Generative Adversarial Network (TS-SRGAN). The network, as shown in Figure~\ref{fig_ecnu}, consists of the pre-Super-Resolution (pre-SR) sub-net for the first stage and a main net for the second stage. Pretrained ADCSR16$\times$~\cite{9022376} is used for the pre-SR sub-net to super-resolve the LR image to pre-SR image with the same resolution as ground truth. The pre-SR sub-net is frozen during the training phase of TS-SRGAN. The main sub-net is composed of a head 3$\times$3 convolutional layer, a de-sub-pixel convolutional layer, a non-linear feature mapping module, an upsampling skip connection, a sub-pixel convolutional layer, a global skip connection and a tail 3$\times$3 convolutional layer.
The Adaptive WDSR-A-Block is modified from WDSR-A-Block \cite{yu2018wide} by adding learnable weight for body scaling and learnable weight for residual scaling. The loss function and the training processes are the same as ESRGAN \cite{wang2018esrgan}.

%\vspace{-0.2cm}

\subsection*{SIA}
\begin{figure}[!htbp]
    \centering
    \subfigure[Upsampling block of generator]{%
    \includegraphics[width=0.8\linewidth]{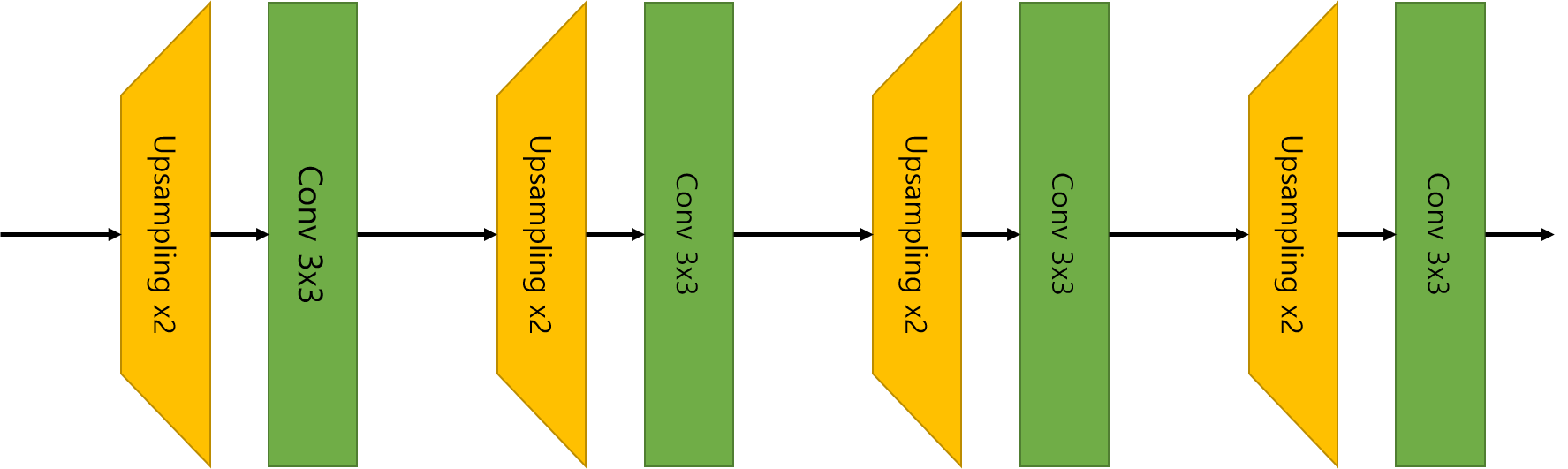}
    }
    \subfigure[Discriminator]{%
    \includegraphics[width=\linewidth]{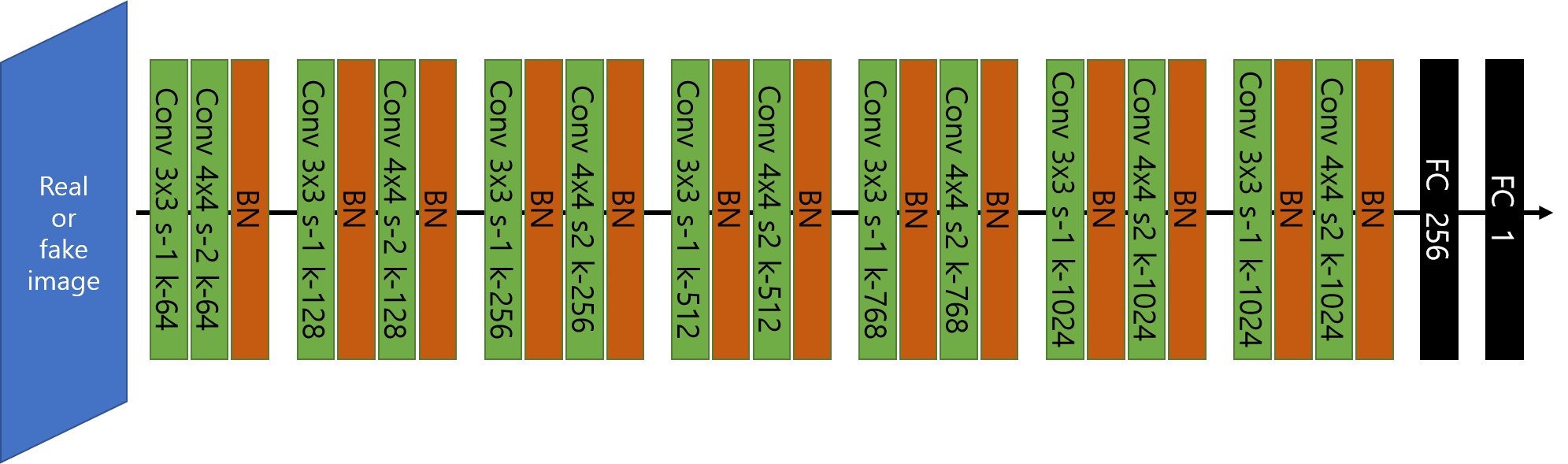}
    }
    %\vspace{-1mm}
    \caption{SIA's network architecture.}
    \label{fig:sia}
\end{figure}

SIA adopted ESRGAN~\cite{wang2018esrgan} to extremely super-resolve an input image with a magnification factor of $16$. The architecture of the generator is almost identical to that of ESRGAN except that the upscaling block of the generator consists of four upsampling layers, each of which doubles the feature size. The discriminator has seven downsampling blocks, each of which consists of two convolutional layers and batch-normalization (BN) layers~\cite{Ioffe-ICML-2015} between them, while the first downsampling block has one BN layer. At the tail of the discriminator, two fully connected (FC) layers with size $256$ and $1$ is applied. See Figure~\ref{fig:sia} for the details.
The hyper-parameters, such as optimizer, learning rate and coefficients for loss terms, are the same as ESRGAN.

%\vspace{-0.2cm}

\subsection*{TTI}
\begin{figure}[!htbp]
    \centering
    \includegraphics[width=\linewidth]{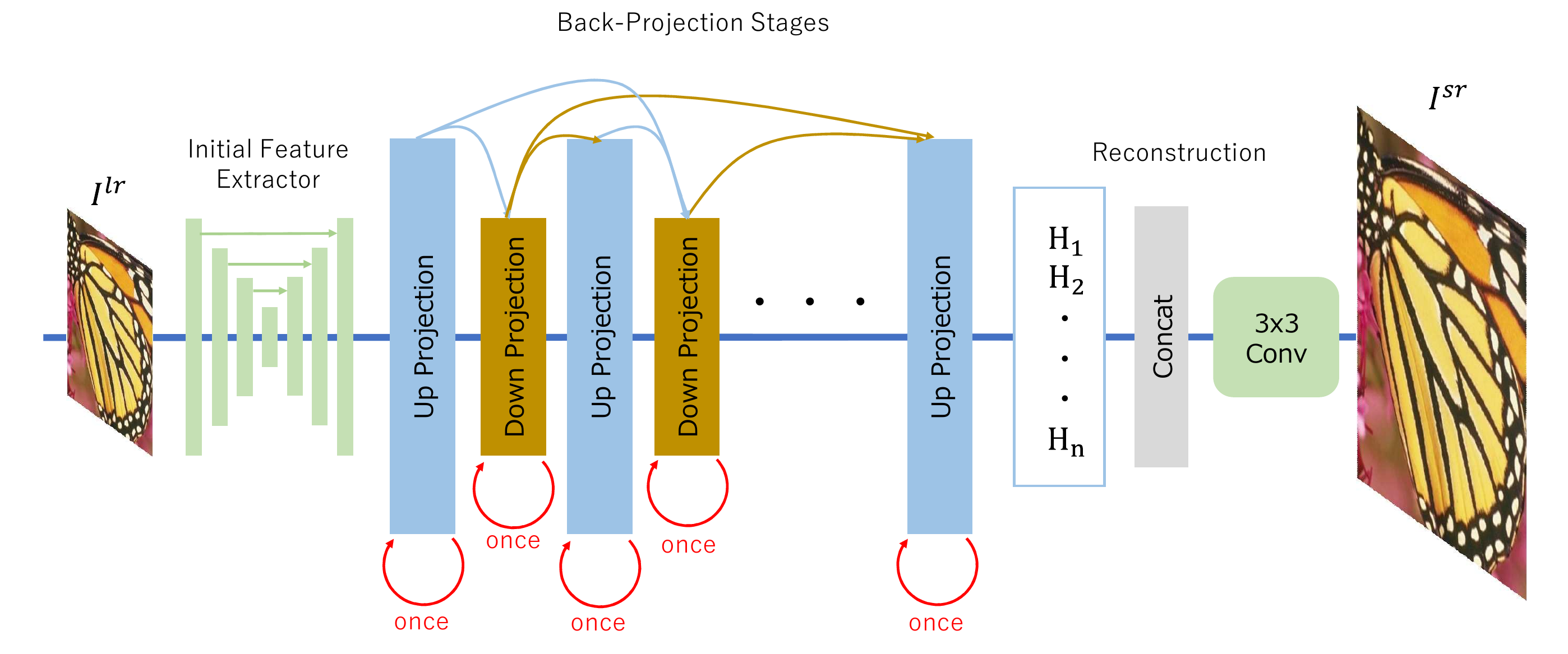}
    \caption{TTI's network architecture.}
    \label{fig:tti}
\end{figure}

\begin{figure}[!htbp]
    \centering
    \includegraphics[width=\linewidth]{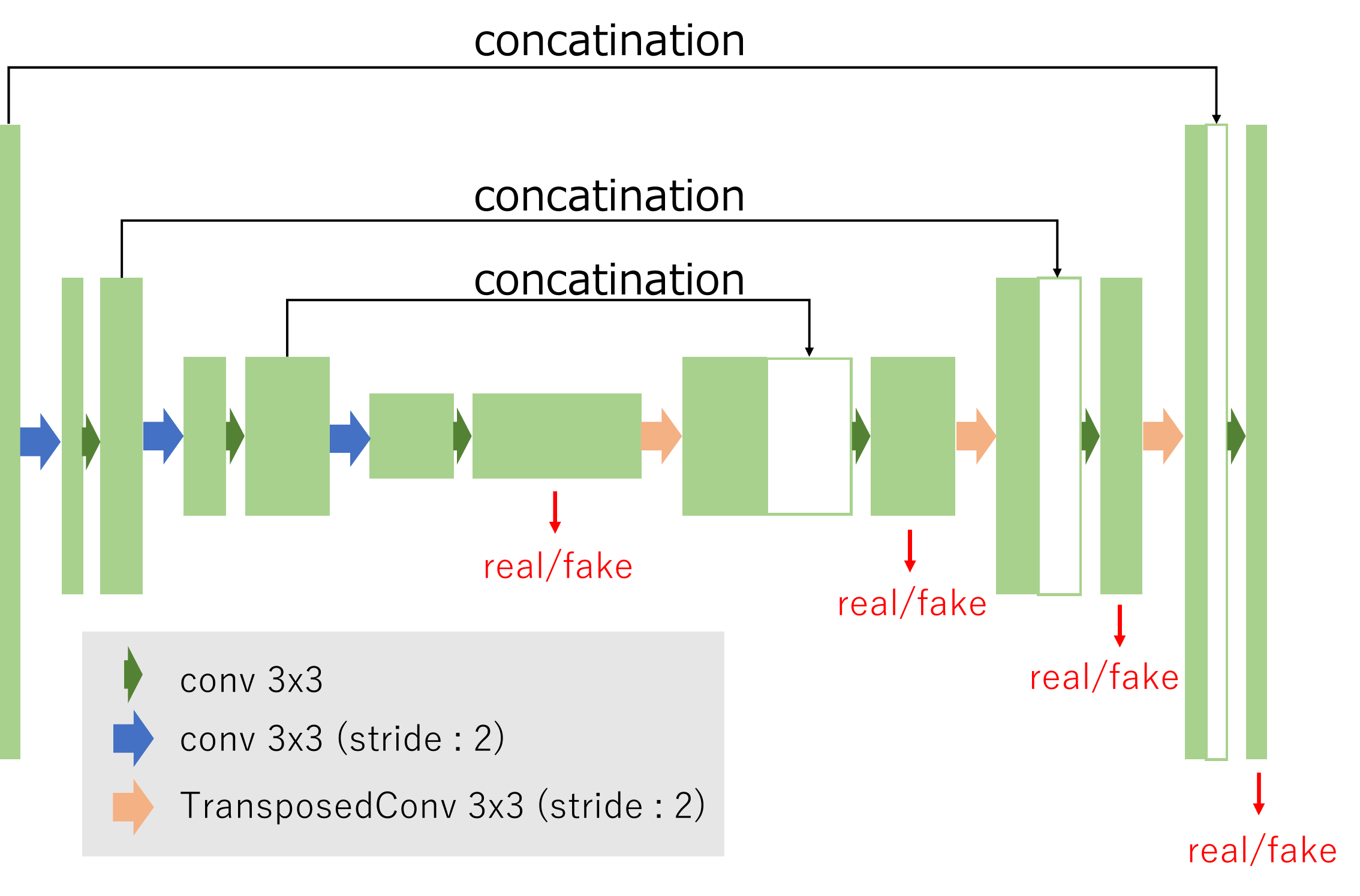}
    \caption{TTI's UNet-like discriminator.}
    \label{fig:disc_arch.eps}
\end{figure}

Inspired by deep back-projection network~\cite{haris2018deep} which introduces iterative up-down projection units for mutual relation between LR and HR feature maps, TTI proposed recurrent progressive perceptual DBPN where each up- and down-projection unit has one $\times$4 and $\times\frac{1}{4}$ scaling layers and performs twice to expand LR images by a scale factor of $\times$16 (see Figure~\ref{fig:tti}). Such a network design not only reduces the number of model parameters but eases the training. For the discriminator, an UNet-like adversarial network (see Figure~\ref{fig:disc_arch.eps}) inspired by~\cite{schonfeld2020u} was used as it can capture both global features (\eg, geometric or structural patterns) and local features (\eg, texture patterns).

%\vspace{-0.2cm}

%\vspace{-0.2cm}

\subsection*{{DeepBlueAI}}

DeepBlueAI proposed bag of tricks for perceptual extreme SR. Based on RCAN~\cite{zhang2018rcan}, various experiments were conducted to explore how to improve PSNR and SSIM. According to the experiments, solutions based RCAN yielded best performance. In the final model, the number of residual groups is 10 and the number of channel in each layer is 128, the number of residual channel attention blocks in each residual group is 20. To obtain the $\times$16 model, a $\times$4 model is first trained from scratch as a pre-trained network after its convergence.

In the training, 1,500 training images with random horizontal flip and rotation are used. In each training batch, 16 LR patches with the size of 48$\times$48 are extracted as inputs. The model is trained by ADAM optimizer with an initial leaning rate $1.0\times10^{-4}$.  The learning rate utilize a cosine annealing schedule~\cite{DBLP:journals/corr/LoshchilovH16a} with total $4.0\times 10^{5}$ iterations and restarts every $1.0\times 10^{5}$ iterations.
Following~\cite{Qiu2019EmbeddedBR}, a $\times$16 model with L1 loss is first trained and then is fine-tuned with L2 loss.
To enhance the PSNR and SSIM, both self-ensemble and model-ensemble are utilized.

\subsection*{APTX4869}
\begin{figure}[!htbp]
    \centering
    \includegraphics[width=\linewidth]{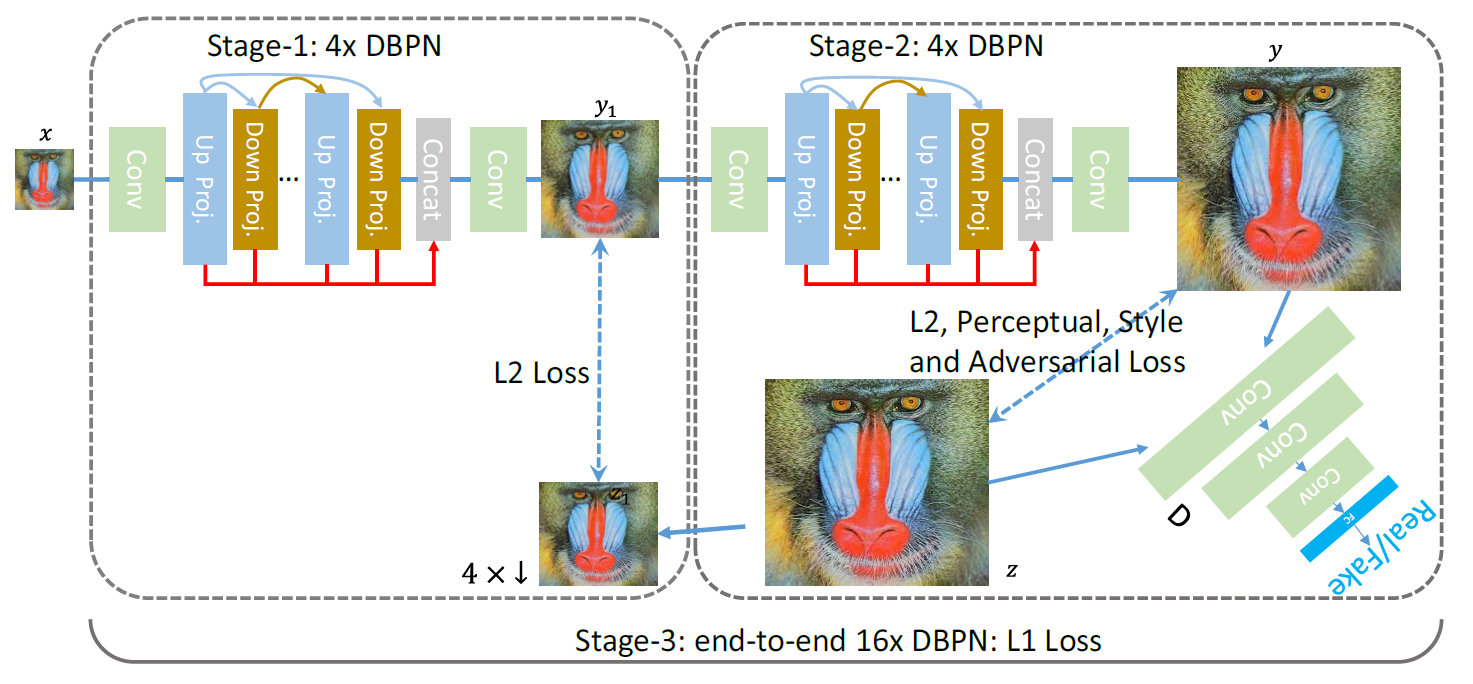}
    %\vspace{-1mm}
    \caption{APTX4869's network architecture.}
    \label{fig:overview}
\end{figure}

APTX4869 proposed progressive super-resolving and refining to tackle the perceptual extreme SR problem. The proposed network, as shown in Figure~\ref{fig:overview}, is decomposed into two cascaded $4\times$ super-resolution ones with three training stages.
In the first stage, the $\times$4 DBPN~\cite{haris2018deep} upsampling sub-module is trained to reconstruct an intermediate artifacts-free result. In the second stage, another $\times$4 upsampling sub-module is then trained with generative adversarial settings to better recover details. In the third stage, the overall architecture is end-to-end refined with L1 loss to eliminate annoying artifacts.

\vspace{-0.2cm}

\subsection*{CNDP-Lab}
\begin{figure}[!htbp]
    \centering
    \includegraphics[width=\linewidth]{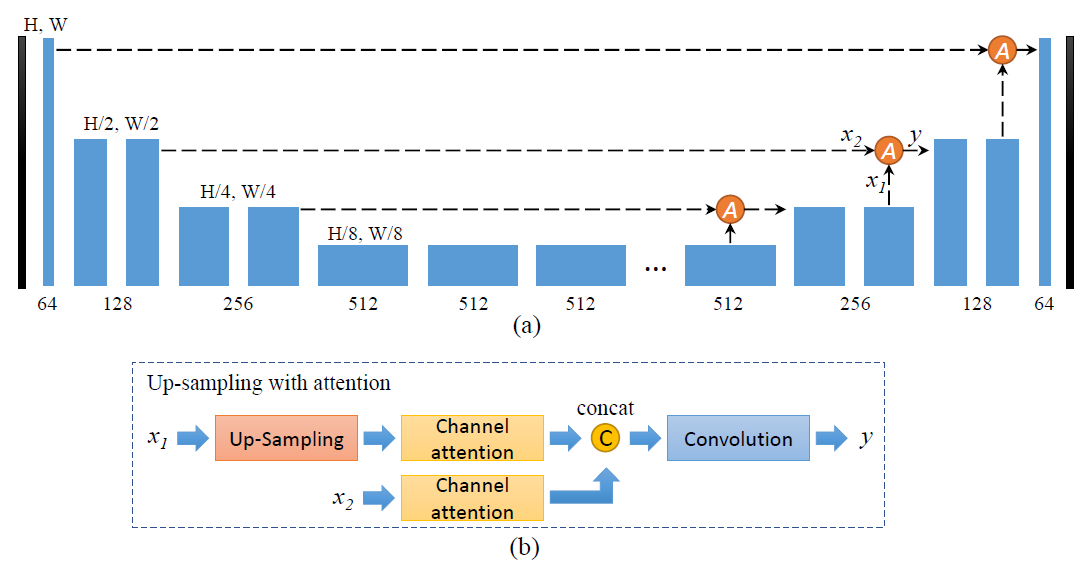}
    %\vspace{-1mm}
    \caption{CNDP-Lab's U-Net with channel attention network architecture.}
    \label{fig:CNDP-Lab}
\end{figure}

CNDP-Lab proposed a cascaded U-Net with channel attention for extreme SR.
The network, as shown in Figure~\ref{fig:CNDP-Lab}, is developed from the traditional U-Net~\cite{ronneberger2015u} by embedding channel attention~\cite{zhang2018rcan} into
the up-sampling process. Before concatenating the two feature maps of $x_1$ and $x_2$,
the features are optimized by channel attention blocks, as shown in Figure~\ref{fig:CNDP-Lab}(b).
The down-sampling process is implemented by a
convolutional layer with a stride of two, and the up-sampling process
adopts the pixel shuffle method.
After each down-sampling or up-sampling layer, a set of residual
channel attention blocks are added to learn features at each scale.
To up-scale the input low-resolution image with a factor of 16, two
U-Net models are trained and each of the model up-scales the input image with a factor
of 4. Because the input and output image of the U-Net have the same resolution,
the input image is first up-scaled with a factor of 4 by bicubic
interpolation.

\vspace{-0.2cm}

\subsection*{MSMers}
The MSMers team proposed PURCAN which adopts progressive upsampling architecture and progressive training strategy for extreme SR. PURCAN takes RCAN~\cite{zhang2018rcan} without the final upsampler as the backbone. The feature maps are progressively upsampled ($\times$1 to $\times$4 to $\times$16) using pixel shuffle. The RCAN backbone is applied into $1\times$ scale while four residual channel attention blocks are applied into the $\times$4 scale to enhance the performance. During training, the backbone part is first trained using $\times$4 dataset which is constructed from DIV8K HR training images. Then the remaining part of the model is added to train on $\times$16 DIV8K training images.

\subsection*{kaws}
\begin{figure}[!htbp]
   \centering
   \includegraphics[width=\linewidth]{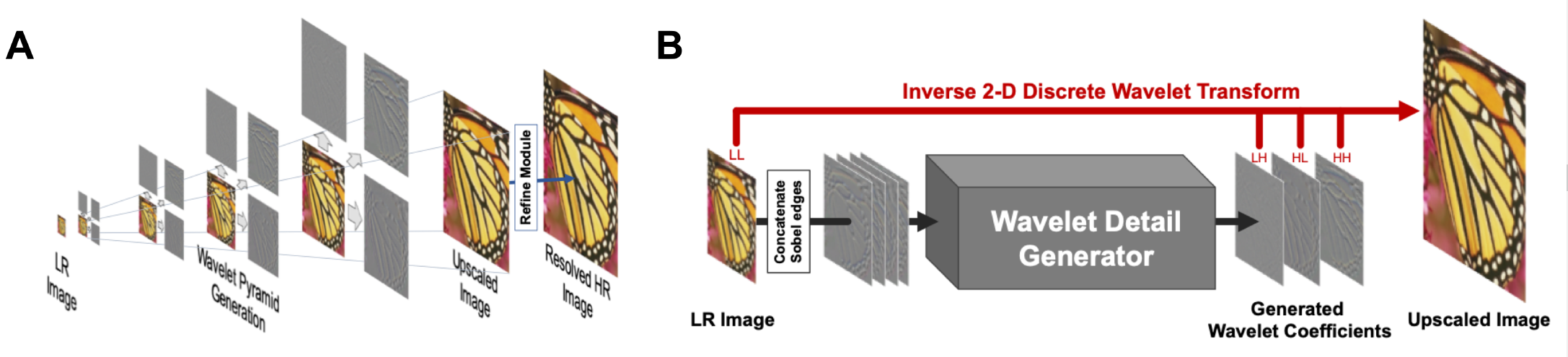}
   %\vspace{-1mm}
   \caption{The network architecture of kaws team.}
   \label{fig:kaws}
\end{figure}

The kaws team proposed wavelet pyramid generation based high-frequency recovery for extreme SR.
In the first stage, wavelet pyramid is generated from low-resolution (LR) image to upscale the image with explicitly recovered high-frequency details.
The upscaled image is then refined from upscaled image to follow the ground-truth HR image in the second stage (see Figure~\ref{fig:kaws}(A)).
To generate the wavelet detail coefficients from the LR image, they first concatenate outputs from convolving the LR image and sobel edge filters of four directions (vertical, horizontal, two-way diagonal).
The concatenated data is input to the wavelet detail generator to output the LH, HL, HH wavelet detail coefficients (see Figure~\ref{fig:kaws}(B)).
The proposed method is expected to have two advantages respect to image quality and generalizability.
In terms of image quality, sharp edges can be recovered by upscaling the image with explicitly generated wavelet detail coefficients, which represent the local high-frequency information of each pixels.
For generalizability, refining the upscaled image can be done with any state-of-the-art super-resolution (SR) model and is expected to improve its performance.
The wavelet detail generator and the refine module are based on the modified U-Net~\cite{ronneberger2015u} and the EDSR~\cite{lim2017enhanced}, respectively.

\vspace{-0.2cm}

\subsection*{UIUC-IFP}
Inspired by progressive multi-scale modelling, UIUC-IFP extended WDSR~\cite{yu2018wide} from single-scale to multi-scale. The proposed
progressive WDSR consists of 4 stages for $\times$16 image SR.
Starting from original feature spatial size, features
are enlarge 2 times spatially and width are reduced 2
times at the end of each stage.

\subsection*{KU\_ISPLB}
\begin{figure}[!htbp]
    \centering
    \includegraphics[width=\linewidth]{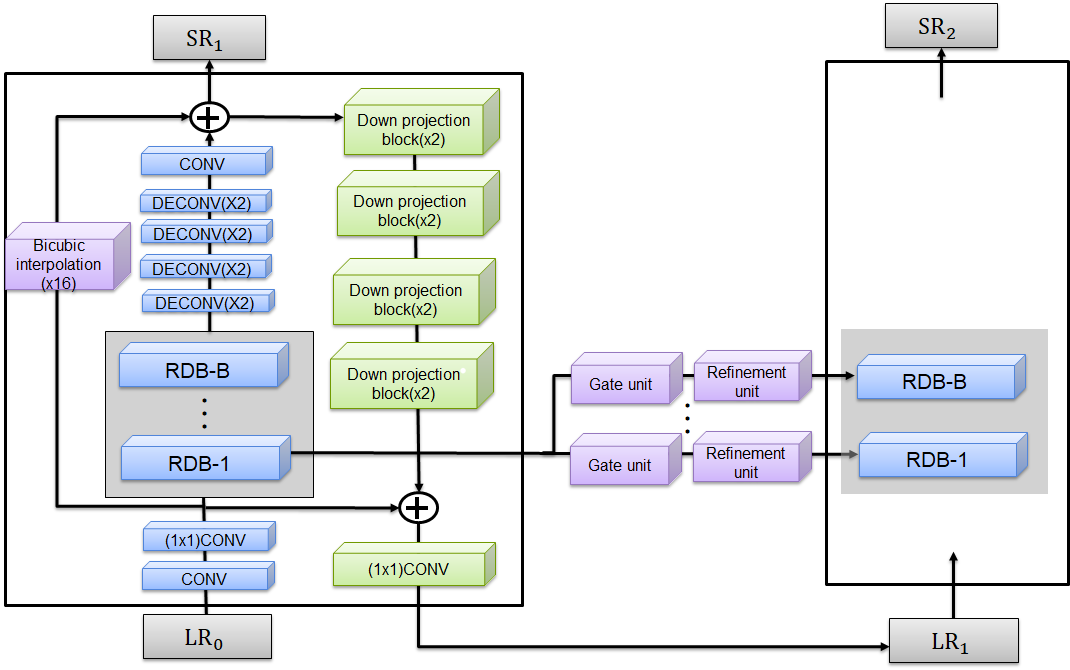}
    \caption{KU\_ISPLB's network architecture.}
   \label{kub}
\end{figure}

KU\_ISPLB proposed feedback recurrent neural network (FBRNN) for extreme SR~\cite{lee2020ntire}. As shown in Figure~\ref{kub}, the LR image is combined with improved LR every step. Inspired by GMFN~\cite{li2019gated} and DPBN~\cite{haris2018deep}, improved LR is given by upsampling network and down-projection network. The the image back projection network is performed 4 times  after the each super resolution network. After yielding better LR image from the HR images in the process, the improved LR is combined with the original LR to prevent bias or over-fitting problems.

\subsection*{MsSrModel}
\begin{figure}[!h]
    \centering
    \includegraphics[width=\linewidth]{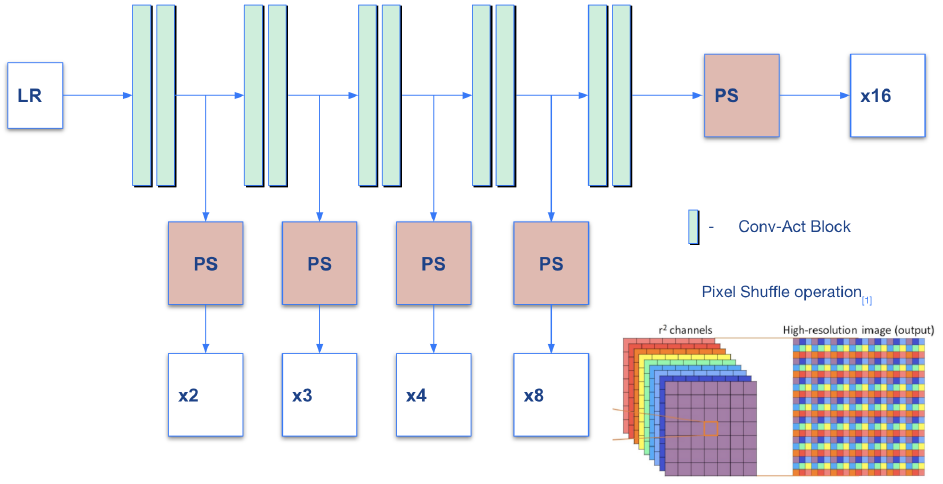}
    \caption{MsSrModel's network architecture.}
    \label{MsSrModel}
\end{figure}

MsSrModel proposed multi-scale SR model (MsSrModel) (see Fig.~\ref{MsSrModel}) for extreme SR. The main idea is to have 5 different optimizations simultaneously for 5 different resolutions.
The model operates on an LR image with pixel shuffle operation~\cite{shi2016real} at the end of each optimization.
L2 loss and VGG perceptual loss are used to optimize the model.

\vspace{-0.2cm}

\subsection*{MoonCloud}
MoonCloud proposed Multi-scale ResNet for perceptual extreme SR. The network uses 16 residual blocks for feature extraction, then adopts a sub-pixel layer to construct the HR features, and finally incorporates a convolution layer with size of 8$\times$3$\times$3$\times$3 to produce the final output.
Based on SRResNet~\cite{ledig2017photo}, a multi-scale strategy is used to learn rich feature for image restoration. In particular, multiple upsampling layers are used to implement multi-scale image SR.

\subsection*{SuperT}
\begin{figure}[!h]
    \centering
   \includegraphics[width=\linewidth]{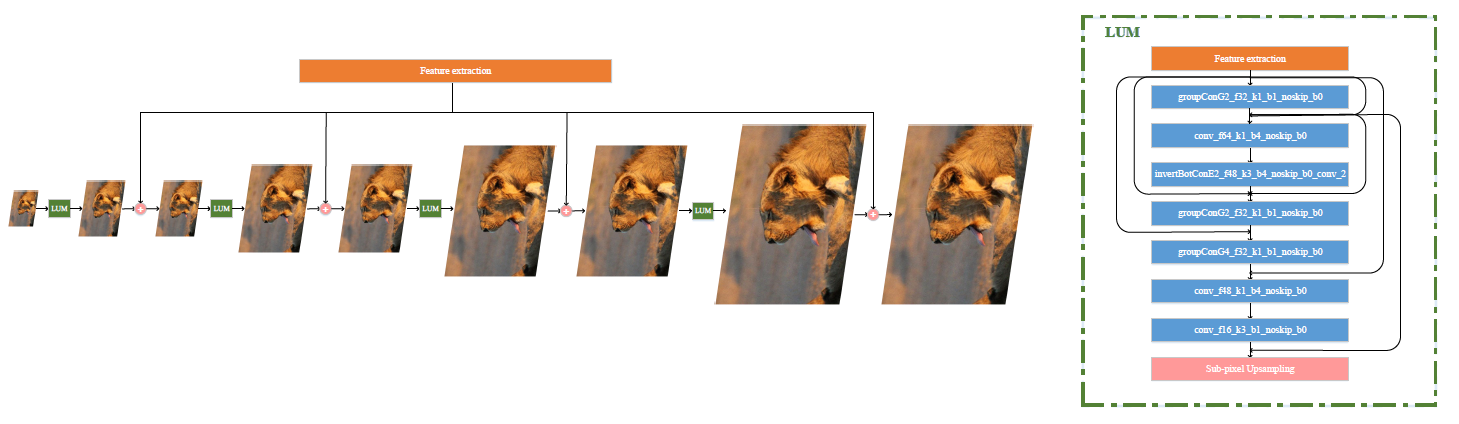}
    %\vspace{-1mm}
   \caption{SuperT's network architecture.}
    \label{fig:team_name}
\end{figure}

SuperT proposed fast and balanced Laplacian pyramid networks for progressive image super-resolution. The network, as shown in Figure~\ref{fig:team_name}, takes LR images as input and gradually predicts sub-band residuals from coarse to fine. At each level, the feature maps are first extracted to reconstruct a higher level image by using a lightweight upsampling module (LUM) with relatively sparse connections. Finally, convolutional layers are adopted to predict subband residuals. The prediction residuals at each level are used to efficiently reconstruct HR images through upsampling and addition operations.

%\vspace{-0.2cm}

\subsection*{sysu-AIR}
\begin{figure}[!htbp]
   \centering
   \includegraphics[width=\linewidth]{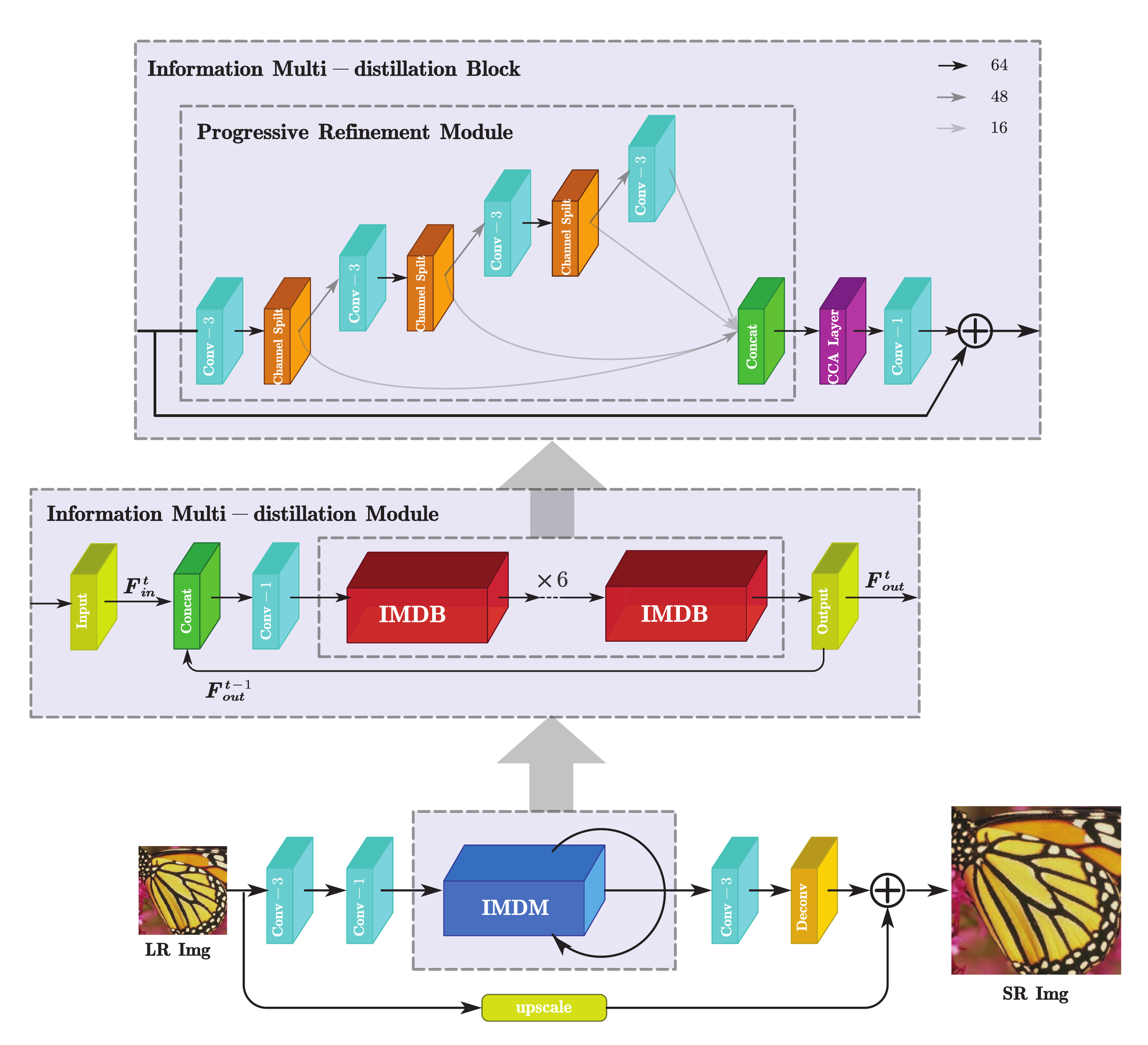}
   %\vspace{-1mm}
   \caption{sysu-AIR's network architecture.}
   \vspace{-0.1in}
   \label{fig:sysu-AIR}
\end{figure}

The sysu-AIR team proposed a fast feedback network for large scale image super-resolution. Inspired by SRFBN~\cite{li2019feedback} and IMDN~\cite{hui2019lightweight}, the proposed Fast-SRFBN is still reserved the RNN structure but with a  information multi-distillation module (IMDM), which can benefit image SR tasks and accelerate inference speed. As shown in Figure~\ref{fig:sysu-AIR}, the IMDM recurrently refines the LR image in a ``coarse to fine'' manner, and it consists of a 1$\times$1 convolutional layer and several stacked information multi-distillation blocks (IMDB).
Benefit from the RNN structure, the proposed network is lightweight but efficient. The final model was trained with L1 loss, VGG perceptual loss, GAN loss, total variation loss, and the novel Fourier spectrum loss~\cite{Zhang19gan}.

\subsection*{{KU\_ISPL\_A}}
\begin{figure}[!htbp]
    \centering
    \includegraphics[width=0.9\linewidth]{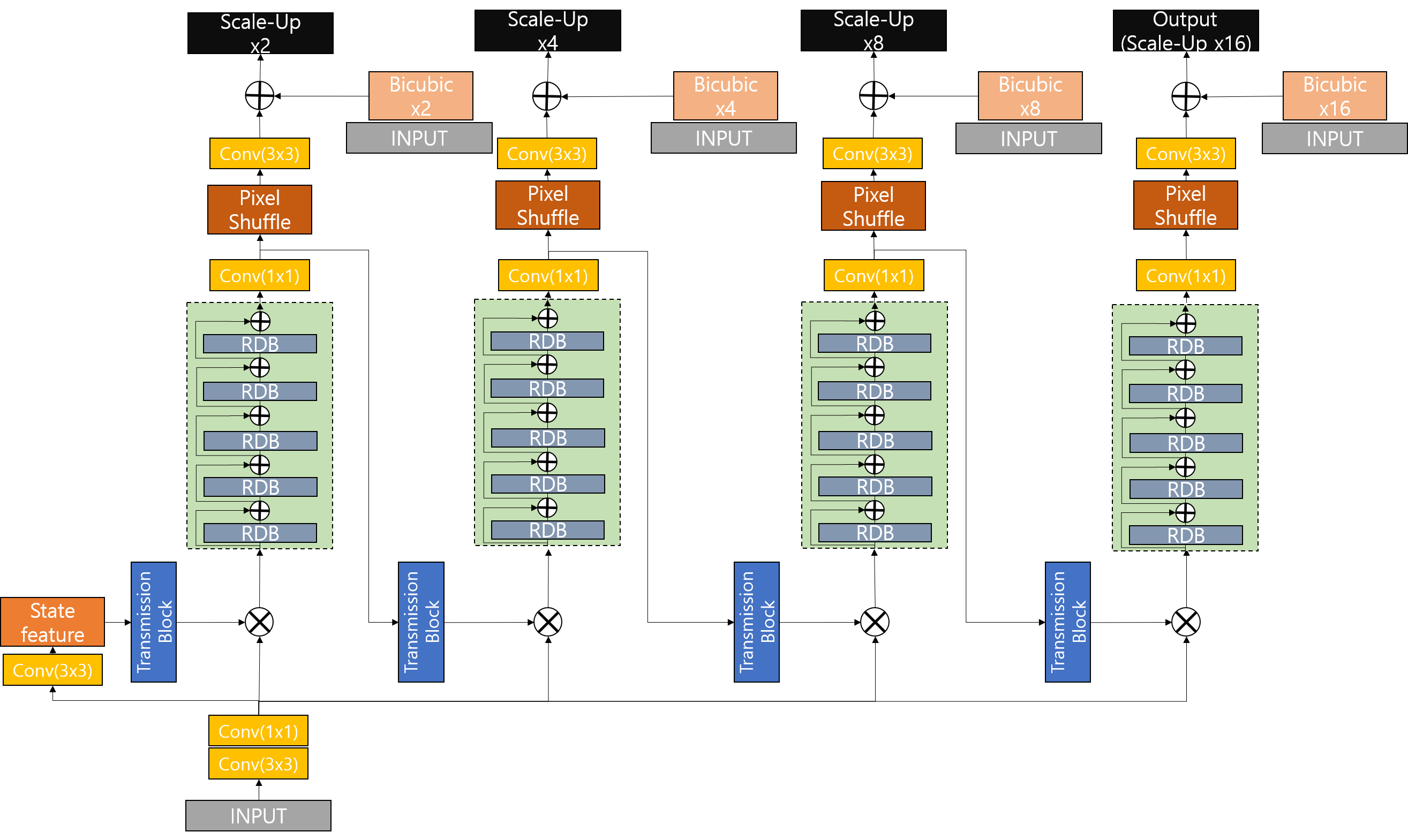}
    \caption{KU\_ISPL\_A's network architecture.}
    \label{fig:RTNet}
\end{figure}

Inspired by SRFBN~\cite{li2019feedback} and GMFN~\cite{li2019gated}, KU\_ISPL\_A proposed a recurrent transmission network (see Figure~\ref{fig:RTNet}) which gradually grows the resolution for each step in the recurrent structure and uses the RDB structure of the RDN~\cite{zhang2018residual} model to extract features for each resolution. Each step produces $\times$2, $\times$4, $\times$8 and $\times$16 resolution results.

\vspace{-0.2cm}

\subsection*{CET\_CVLab}
\begin{figure}[!htbp]
  \centering
  \includegraphics[width=\linewidth]{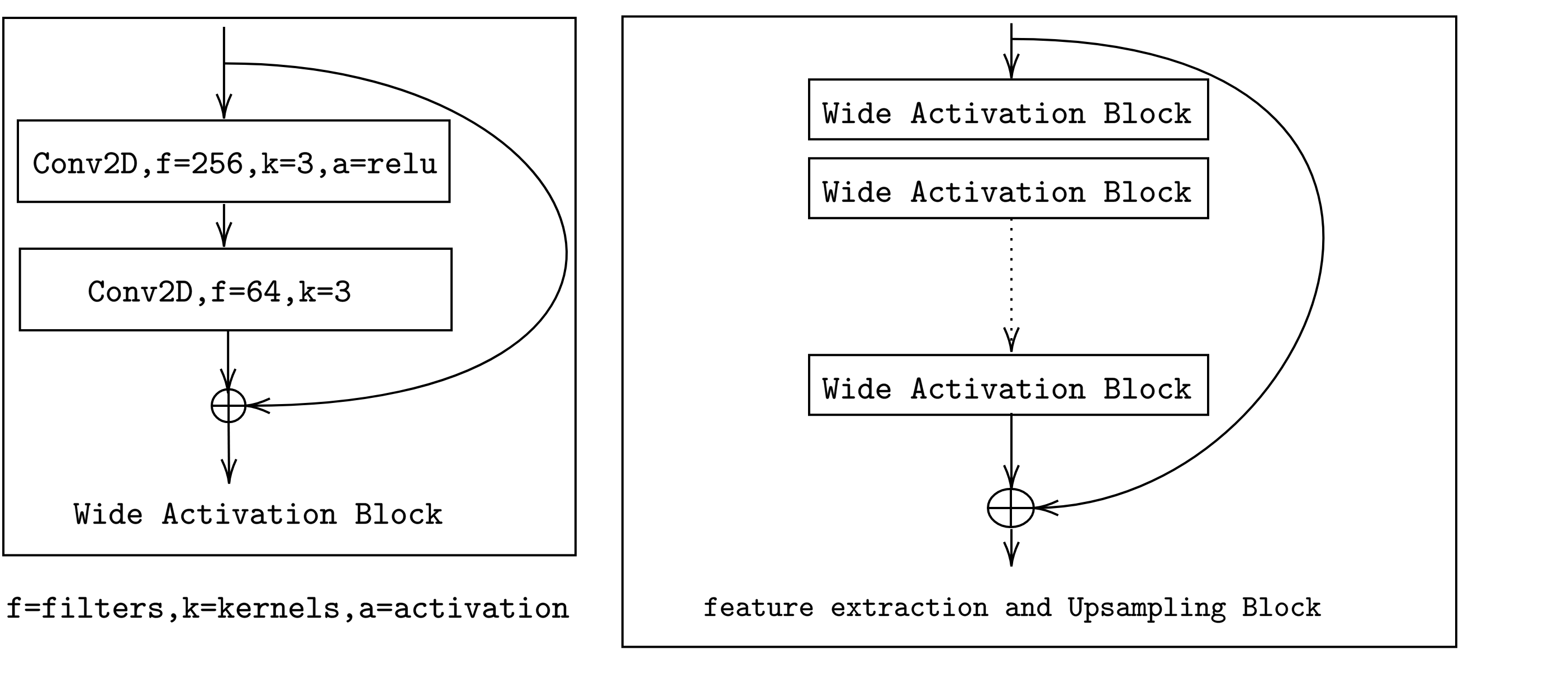}
  \includegraphics[width=\linewidth]{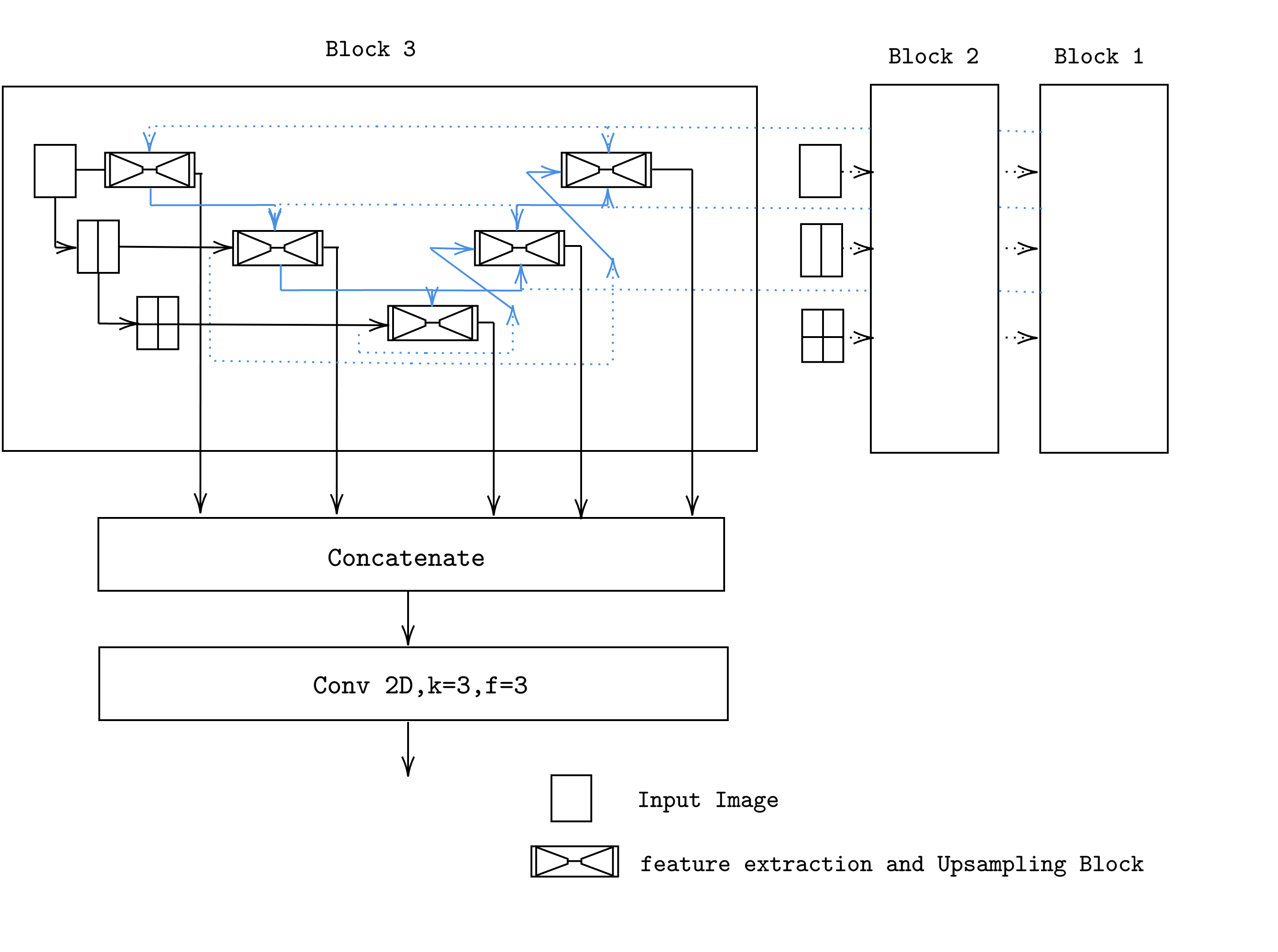}
  \caption{CET\textunderscore CVLab's network architecture.}
 \label{fig:v-stacked}
\end{figure}

CET\_CVLab proposed a V-Stacked relativistic GAN for extreme SR. In the generator part of the network, there are 3 stacks of deep CNN based structure which is inspired from \cite{v_stacked}. Within each stack there is a pyramidical arrangement of layers that forms the V-shape as shown in Figure~\ref{fig:v-stacked}. Each level of the structure is an encoder-decoder block. There are 5 levels in the pyramid and the initial level takes the whole image as a single patch followed by layers with patches up to 4 and then back to a single patch. All these image patches are passed through a feature extraction network followed by $\times$16 upsampling layer. The generator network is first trained with L1 loss, and then refined with relativistic average GAN loss and VGG perceptual loss.

\vspace{-0.2cm}

\section*{Acknowledgements}
We thank the NTIRE 2020 sponsors: HUAWEI, OPPO, Voyage81, MediaTek, DisneyResearch$\vert$Studios, and Computer Vision Lab (CVL) ETH Zurich.

%------------------------------------------------------------------------
\appendix
\section{Teams and affiliations}
\label{sec:teams}

\subsection*{NTIRE2020 team}
\noindent\textit{\textbf{Title: }} NTIRE 2020 Perceptual Extreme Super-Resolution Challenge\\
\noindent\textit{\textbf{Members: }} \\
Kai Zhang (\href{mailto:kai.zhang@vision.ee.ethz.ch}{kai.zhang@vision.ee.ethz.ch}),\\
Shuhang Gu (\href{mailto:shuhang.gu@vision.ee.ethz.ch}{shuhang.gu@vision.ee.ethz.ch}),\\
Radu Timofte (\href{mailto:radu.timofte@vision.ee.ethz.ch}{radu.timofte@vision.ee.ethz.ch})\\
\noindent\textit{\textbf{Affiliations: }}\\
Computer Vision Lab, ETH Zurich, Switzerland\\

\subsection*{OPPO-Research}
\noindent\textit{\textbf{Title: }}Perceptual Extreme Super-Resolution Network with Receptive Field Block\\
\noindent\textit{\textbf{Members: }}\textit{Taizhang Shang
\\\noindent(\href{mailto:username@mail.com}{shangtaizhang@oppo.com)}}, Qiuju Dai, Shengchen Zhu, Tong Yang, Yandong Guo\\
\noindent\textit{\textbf{Affiliation: }}\\
OPPO Research\\

\vspace{-0.4cm}

\subsection*{CIPLAB}
\noindent\textit{\textbf{Title: }}Investigating Loss Functions for Extreme Super-Resolution\\
\noindent\textit{\textbf{Members: }}\textit{Younghyun Jo$^{1}$
\\\noindent(\href{mailto:yh.jo@yonsei.ac.kr}{yh.jo@yonsei.ac.kr}}), Sejong Yang$^{1}$, Seon Joo Kim$^{1,2}$\\
\noindent\textit{\textbf{Affiliation: }}\\
$^{1}$ Yonsei University\\
$^{2}$ Facebook\\
\vspace{-0.4cm}

\subsection*{HiImageTeam}
\noindent\textit{\textbf{Title: }}Cascade SR-GAN for Extreme Super-Resolution\\
\noindent\textit{\textbf{Members: }}\textit{Lin Zha$^{1}$
\\\noindent(\href{mailto:zhalin@hisense.com}{zhalin@hisense.com)}}, Jiande Jiang$^{1}$, Xinbo Gao$^{2}$, Wen Lu$^{2}$\\
\noindent\textit{\textbf{Affiliation: }}\\
$^{1}$ Qingdao Hi-image Technologies Co.,Ltd (Hisense Visual Technology Co.,Ltd.)\\
$^{2}$ Xidian University
\\

\vspace{-0.4cm}

\subsection*{ECNU}%Team Name
\noindent\textit{\textbf{Title: }}Two-Stages-SRGAN\\
\noindent\textit{\textbf{Members: }}\textit{Jing Liu
\\\noindent(\href{51174500035@stu.ecnu.edu.cn}{51174500035@stu.ecnu.edu.cn})}\\
\noindent\textit{\textbf{Affiliation: }}\\
Multimedia and Computer Vision Lab, East China Normal University (ECNU)\\

\vspace{-0.4cm}

\subsection*{SIA}
\noindent\textit{\textbf{Title: }}Perception-Oriented Extreme Upscaling using ESRGAN\\
\noindent\textit{\textbf{Members: }}\textit{Kwangjin Yoon
\\\noindent(\href{mailto:yoon28@si-analytics.ai}{yoon28@si-analytics.ai})}, Taegyun Jeon\\
\noindent\textit{\textbf{Affiliation: }}\\
SI Analytics Co., Ltd., 441 Expo-ro, Yuseong-gu, Daejeon, 34051, Republic of Korea\\

\vspace{-0.4cm}

\subsection*{TTI}
\noindent\textit{\textbf{Title: }}Deep Back Projection for Perceptual Extreme Super-Resolution\\
\noindent\textit{\textbf{Members: }}\textit{Kazutoshi Akita
\\\noindent(\href{mailto:sd19401@toyota-ti.ac.jp}{sd19401@toyota-ti.ac.jp})}, Takeru Ooba, Norimichi Ukita\\
\noindent\textit{\textbf{Affiliation: }}\\
Toyota Technological Institute (TTI)\\

\vspace{-0.4cm}

\subsection*{DeepBlueAI}
\noindent\textit{\textbf{Title}}: Bag of Tricks for Perceptual Extreme Super-Resolution
\noindent\textit{\textbf{Members: }}\textit{Zhipeng Luo
\\\noindent(\href{mailto:luozp@deepblueai.com}{luozp@deepblueai.com})}, Yuehan Yao, Zhenyu Xu\\
\noindent\textit{\textbf{Affiliation: }}\\
DeepBlue Technology (Shanghai) Co.,Ltd\\

\vspace{-0.4cm}

\subsection*{APTX4869}
\noindent\textit{\textbf{Title: }}Progressive Super-Resolving and Refining\\
\noindent\textit{\textbf{Members: }}\textit{Dongliang He$^{1}$
\\\noindent(\href{mailto:hedlcc@126.com}{hedlcc@126.com})}, Wenhao Wu$^{2}$, Yukang Ding$^{1}$, Chao Li$^{1}$, Fu Li$^{1}$, Shilei Wen$^{1}$\\
\noindent\textit{\textbf{Affiliation: }}\\
$^{1}$ Department of Computer Vision Technology (VIS), Baidu Inc.\\
$^{2}$ Shenzhen Institutes of Advanced Technology, Chinese Academy of Sciences, China.
\\

\vspace{-0.4cm}

\subsection*{CNDP-Lab}
\noindent\textit{\textbf{Title: }}Cascaded U-Net with Channel Attention for Image Super-Resolution\\
\noindent\textit{\textbf{Members: }}\textit{Jianwei Li$^{1,2}$
\\\noindent(\href{mailto:ljwdust@gmail.com}{ljwdust@gmail.com)}}\\
\noindent\textit{\textbf{Affiliation: }}\\
$^{1}$ Peking University\\
$^{2}$ State Key Laboratory of Digital Publishing Technology, Founder Group
\\

\vspace{-0.4cm}

\subsection*{MSMers}
\noindent\textit{\textbf{Title: }}{Progressively Upsampled Residual Channel Attention Network for Extreme Super-Resolution}\\
\noindent\textit{\textbf{Members: }}Fuzhi Yang$^{1}$
(\href{mailto:yfzcopy0702@sjtu.edu.cn}{yfzcopy0702@sjtu.edu.cn}), Huan Yang$^{2}$(\href{mailto:huayan@microsoft.com}{huayan@microsoft.com}), Jianlong Fu$^{2}$\\
\noindent\textit{\textbf{Affiliation: }}\\
$^{1}$ Shanghai Jiao Tong University\\
$^{2}$ Microsoft Research, Beijing, P.R. China\\

\vspace{-0.4cm}

\subsection*{kaws}
\noindent\textit{\textbf{Title: }}Wavelet Pyramid Generation based High-frequency Recovery for Perceptual Extreme Super-Resolution\\
\noindent\textit{\textbf{Members: }}\textit{Byung-Hoon Kim$^{1}$
\\\noindent(\href{mailto:egyptdj@kaist.ac.kr}{egyptdj@kaist.ac.kr})}, JaeHyun Baek$^{2}$, Jong Chul Ye$^{1}$\\
\noindent\textit{\textbf{Affiliation: }}\\
$^{1}$ Korea Advanced Institute of Science and Technology (KAIST)\\
$^{2}$ Amazon Web Services
\\

\vspace{-0.4cm}

\subsection*{UIUC-IFP}
\noindent\textit{\textbf{Title: }} Progressive WDSR\\
\noindent\textit{\textbf{Members:}}\textit{ Yuchen Fan
\\\noindent(\href{mailto:yuchenf4@illinois.edu}{yuchenf4@illinois.edu})}, Thomas S. Huang\\
\noindent\textit{\textbf{Affiliation: }}\\
University of Illinois at Urbana-Champaign\\

\vspace{-0.4cm}

\subsection*{KU\_ISPLB}
\noindent\textit{\textbf{Title: }} FBRNN:feedback recurrent neural network\\
\noindent\textit{\textbf{Members: }}\textit{Junyeop Lee
\\\noindent(\href{mailto:jylee@ispl.korea.ac.kr}{jylee@ispl.korea.ac.kr})}, Bokyeung Lee, Jungki Min, Gwantae Kim, Kanghyu Lee, Jaihyun Park\\
\noindent\textit{\textbf{Affiliation: }}\\
Korea University\\

\vspace{-0.4cm}

\subsection*{MsSrModel}
\noindent\textit{\textbf{Title: }} Multi-scale SR Model\\
\noindent\textit{\textbf{Members:}}\textit{ Mykola Mykhailych
\\\noindent(\href{mailto:mykolam@wix.com}{mykolam@wix.com})}\\
\noindent\textit{\textbf{Affiliation: }}\\
Wix.com Ltd.\\

\vspace{-0.4cm}

\subsection*{MoonCloud}
\noindent\textit{\textbf{Title: }}Multi-scale ResNet\\
\noindent\textit{\textbf{Members: }}\textit{Haoyu Zhong$^{1}$
\\\noindent(\href{mailto:username@mail.com}{hy0421@outlook.com})}, Yukai Shi$^{1}$, Xiaojun Yang$^{1}$, Zhijing Yang$^{1}$, Liang Lin$^{2}$\\
\noindent\textit{\textbf{Affiliation: }}\\
$^{1}$ Guangdong University of Technology\\
$^{2}$ Sun Yat-sen University
\\

\vspace{-0.4cm}

\subsection*{SuperT}
\noindent\textit{\textbf{Title: }}Fast and Balanced Laplacian Pyramid Networks for Progressive Image Super-Resolution\\
\noindent\textit{\textbf{Members: }}\textit{Tongtong Zhao
\\\noindent(\href{daitoutiere@gmail.com}{daitoutiere@gmail.com)}}, Jinjia Peng, Huibing Wang\\
\noindent\textit{\textbf{Affiliation: }}\\
Dalian Maritime University\\

\vspace{-0.4cm}

\subsection*{sysu-AIR}
\noindent\textit{\textbf{Title: }} A Fast Feedback Network for Large Scale Image Super-Resolution\\
\noindent\textit{\textbf{Members: }}\textit{Zhi Jin
\\\noindent(\href{mailto:jinzh26@mail.sysu.edu.cn}{jinzh26@mail.sysu.edu.cn})},  Jiahao Wu, Yifu Chen, Chenming Shang, Huanrong Zhang\\
\noindent\textit{\textbf{Affiliation: }}\\
School of Intelligent Systems Engineering, Sun Yat-sen University.\\

\vspace{-0.4cm}

\subsection*{KU\_ISPL\_A}
\noindent\textit{\textbf{Title: }}Recurrent Transmission Network for Extreme Super Resolution\\
\noindent\textit{\textbf{Members: }}\textit{Jeongki Min
\noindent(\href{mailto:jkmin@ispl.korea.ac.kr}{jkmin@ispl.korea.ac.kr})},\\ Junyeop Lee, Bokyeung Lee, Jaihyun Park, Gwantae Kim, Kanghyu Lee\\
\noindent\textit{\textbf{Affiliation: }}  Korea University\\

\vspace{-0.4cm}

\subsection*{CET\textunderscore CVLab}
\noindent\textit{\textbf{Title: }}Perceptual Extreme Super resolution Using V-Stacked Relativistic GAN\\
\noindent\textit{\textbf{Members: }}\textit{Hrishikesh P S
\\\noindent(\href{mailto:hrishikeshps@cet.ac.in}{hrishikeshps@cet.ac.in})}, Densen Puthussery, Jiji C V\\
\noindent\textit{\textbf{Affiliation: }}\\
College of Engineering Trivandrum\\

\vspace{-0.4cm}

{\small
\bibliographystyle{ieee_fullname}
\bibliography{egbib}
}

\end{document}